\documentclass{jpp}
\usepackage{graphicx}

\usepackage[utf8]{inputenc}
\usepackage[T1]{fontenc}
\usepackage{amsmath}

\usepackage{mhchem}
\usepackage{gensymb}

\usepackage{xcolor}

\newcommand{\diff}{\mathrm{d}}
\newcommand{\f}{\frac}
\newcommand{\ov}{\overline}
\newcommand{\ii}{\mathrm{i}}

\shorttitle{Multi-dimensional current-carrying instabilities and turbulence}
\shortauthor{W. H. R. Chan, K. Hara and I. D. Boyd}

\title{Effects of multi-dimensionality and energy exchange on electrostatic current-driven plasma instabilities and turbulence}

\author{Wai Hong Ronald Chan\aff{1,2}
  \corresp{\email{Ronald\_Chan@ihpc.a-star.edu.sg}},
  Kentaro Hara\aff{3}
 \and Iain D. Boyd\aff{1}}

\affiliation{\aff{1}Department of Aerospace Engineering Sciences, University of Colorado,
Boulder, CO 80309, USA
\aff{2}Institute of High Performance Computing (IHPC), Agency for Science, Technology and Research (A*STAR), 1 Fusionopolis Way, \#16-16 Connexis, Singapore 138632, Republic of Singapore
\aff{3}Department of Aeronautics and Astronautics, Stanford University, Stanford, CA 94305, USA}

\begin{document}

\maketitle

\begin{abstract}
Large-amplitude current-driven plasma instabilities, which can transition to the Buneman instability, were observed in one-dimensional (1D) simulations to generate high-energy backstreaming ions. We investigate the saturation of multi-dimensional plasma instabilities and its effects on energetic ion formation. Such ions directly impact spacecraft thruster lifetimes and are associated with magnetic reconnection and cosmic ray inception. An Eulerian Vlasov--Poisson solver employing the grid-based direct kinetic method is used to study the growth and saturation of 2D2V collisionless, electrostatic current-driven instabilities spanning two dimensions each in the configuration (D) and velocity (V) spaces supporting ion and electron phase-space transport. Four stages characterise the electric potential evolution in such instabilities: linear modal growth, harmonic growth, accelerated growth via quasi-linear mechanisms alongside non-linear fill-in, and saturated turbulence. Its transition and isotropisation process bears considerable similarities to the development of hydrodynamic turbulence. While a tendency to isotropy is observed in the plasma waves, followed by electron and then ion phase space after several ion-acoustic periods, the formation of energetic backstreaming ions is more limited in the 2D2V than in the 1D1V simulations. Plasma waves formed by two-dimensional electrostatic kinetic instabilities can propagate in the direction perpendicular to the net electron drift. Thus, large-amplitude multi-dimensional waves generate high-energy transverse-streaming ions and eventually limit energetic backward-streaming ions along the longitudinal direction. The multi-dimensional study sheds light on interactions between longitudinal and transverse electrostatic plasma instabilities, as well as fundamental characteristics of the inception and sustenance of unmagnetised plasma turbulence.
\end{abstract}

\section{Introduction}\label{sec:intro}

\setcitestyle{notesep={ }}
Two categories of electrostatic current-carrying plasma instabilities are typically considered~\citep{omura2003particle, mikellides2005hollow}. The current-carrying ion-acoustic instability manifests when the net electron drift speed $\tilde{U}_d=\left|\tilde{U}_e - \tilde{U}_i\right|$ exceeds the ion-acoustic speed $\sqrt{k_\text{B} \tilde{T}_e/\tilde{m}_i}$ and $\tilde{T}_e \gg \tilde{T}_i$, where $\tilde{U}_e$, $\tilde{U}_i$, $k_\text{B}$, $\tilde{T}_e$, $\tilde{T}_i$, and $\tilde{m}_i$ are the electron bulk velocity, ion bulk velocity, Boltzmann constant, electron temperature, ion temperature, and ion mass, respectively, and the tilde notation over problem parameters denotes dimensional quantities~\citep[and references therein]{stringer1964electrostatic}. Over a broad range of $\tilde{T}_e/\tilde{T}_i$, the Buneman instability arises when the net electron drift speed exceeds the electron thermal speed. In a 1D configuration, the threshold is $\tilde{U}_d \gtrsim 1.3 \tilde{c}_e$, where $\tilde{c}_e = \sqrt{k_\text{B} \tilde{T}_e/\tilde{m}_e}$ and $\tilde{m}_e$ is the electron mass~\citep{buneman1959dissipation}. Depending on $\tilde{U}_d$, both instabilities can be physically relevant in the same plasma configuration. In such instabilities, a sufficiently fast net electron drift triggers plasma wave growth. At a significant wave amplitude, a transition to multi-scale electric potential structures occurs, which in turn induces localised ion acceleration with broadband phase-space structures of an ostensibly turbulent nature.
\setcitestyle{notesep={, }}

\setcitestyle{notesep={; }}
An example in which multi-dimensional current-carrying instabilities play an important role is the plume of a hollow cathode, e.g., for electric spacecraft thrusters, surface processing, and plasma discharges~\citep[and references therein]{oks1999development,becker2006microplasmas,goebel2021plasma}. The erosion of cathode structures can limit the lifetime of years-long outer-space missions by an order of magnitude to less than $O(10^4)$ hours~\citep[e.g.,][]{friedly1992high, kameyama2000measurements, williams2000laser, mikellides2005hollow, mikellides2007evidence, mikellides2008wearII, mikellides2015numerical, goebel2007potential, jorns2014ion, lev2019recent}. A key cause of cathode sputtering is surface collisions of energetic ions at its orifice with energies corresponding to up to $100\text{ eV}$~\citep{friedly1992high, williams1992electron, kameyama2000measurements, williams2000laser, boyd2004modeling, goebel2007potential, mikellides2008wearII, farnell2011comparison}. Such high-energy ions are currently postulated to arise in part from the aforementioned collisionless, electrostatic current-carrying instabilities~\citep{williams2000laser, mikellides2005hollow, mikellides2007evidence, mikellides2008wearII, mikellides2015numerical, goebel2007potential, jorns2014ion, lopezortega2016importance, jorns2017propagation, sary2017hollowI, sary2017hollowII, hara2018test, lopezortega2018hollow, hara2019overview}. In particular, the presence of transversely energetic ions has been observed experimentally~\citep{boyd2004modeling, goebel2007potential, farnell2011comparison, hall2019effect}, and recent laser Thomson scattering measurements show evidence of bi-directional plasma waves at oblique orientations to the local magnetic field (and uni-directional along the parallel direction) in the plume of a hollow cathode~\citep{tsikata2021characterization}. Despite the increasing experimental evidence of multi-dimensional electrostatic plasma waves amplified due to kinetic instabilities, there have not been many detailed numerical studies simulating and analysing the long-term behaviour of such waves to date. Characterising these precursors of the erosion process can improve predictions of thruster lifetime and complement accelerated life testing. Additionally, high-energy ions that are accelerated in the longitudinal and transverse directions can be relevant to the inception of magnetic reconnection and its resulting auroral emissions, where high-frequency turbulence and subsequently intense currents are formed during changes in the magnetic topology~\citep[and references therein]{quon1976formation,mozer1980satellite,temerin1982observations,lysak1990electrodynamic,cairns1995electrostatic,ergun1998fast,drake2003formation}.
\setcitestyle{notesep={, }}

The generation of axially energetic ions has been numerically investigated through Eulerian Vlasov--Poisson simulations employing the grid-based direct kinetic (DK) method in a single spatial dimension (1D)~\citep{hara2019ion, vazsonyi2020non}. In this paper, we extend the analysis to two spatial dimensions (2D) to investigate how ions are energised in the transverse (radial) direction. More broadly, we seek a fundamental understanding of the primary mechanisms driving the growth and saturation of multi-dimensional plasma waves driven by current-carrying instabilities, as well as the resulting ion and electron velocity distributions after the system reaches a state of saturated turbulence. Such a study involves simultaneous analysis of the transition pathway to turbulence of the accompanying electric field and then the ion phase-space distribution. We build on previous numerical studies of the 2D Buneman instability~\citep{amano2009nonlinear} but focus on late ion acceleration instead of early electron acceleration and use a physical mass ratio close to that of a hydrogen plasma ($\tilde{m}_i/\tilde{m}_e = 1.8229\times10^3$). Following the pioneering work by~\citet{forslund1970numerical} on electrostatic counterstreaming instabilities, \citet{chapman2021nonlinear} investigated the non-linear evolution of the ion--ion streaming instability for warm electrons and counter-streaming ions with streaming speeds on the order of the ion-acoustic speed. We analyse the growth and saturation characteristics of 2D current-carrying instabilities for an ion--electron system with larger net drift speeds on the order of the electron thermal speed, focusing on plasma wave growth due to the presence of the net current. In addition, we take a deeper dive into the key stages underlying the genesis of multi-dimensional electrostatic plasma turbulence, and examine their implications on energetic ion formation and hollow cathode sputtering. Several $ \tilde{U}_d/\tilde{c}_e$ ratios are considered to span possible cathode operating conditions. A number of 1D and 2D instability characteristics are also compared to establish connections with previous 1D work and elucidate the role of multi-dimensionality. In this work, we consider $\tilde{T}_e/\tilde{T}_i = 10$, which is representative of cathode operating conditions~\citep{goebel2005hollow, mikellides2005hollow, mikellides2007evidence, mikellides2008wearII, farnell2011comparison, jorns2017propagation}.

The objective of this work is to analyse the stages of collisionless, electrostatic current-carrying instability growth, as well as the inception of multi-dimensional electrostatic plasma turbulence and its spectral characteristics. In \S~\ref{sec:setup}, we describe our computational and physical set-up. In \S~\ref{sec:2D}, we discuss results of the 2D instability. These are compared with key results from the 1D instability in \S~\ref{sec:1D}. Conclusions are provided in \S~\ref{sec:conc}.

\section{Methodology}\label{sec:setup}

\setcitestyle{notesep={; }}
While the particle-in-cell method is the predominant kinetic simulation approach, such particle-based kinetic methods are susceptible to statistical noise. This is because a sufficiently large number of super-particles is required in each computational cell, together with sufficiently long time averaging for steady and quasi-steady configurations, to adequately sample the particle distribution function in the configuration and velocity spaces~\citep{chen1996statistical,kannenberg2000strategies,farbar2010modeling}. This issue is particularly exacerbated in current-carrying instabilities whose phase velocity coincides with the tails of the ion velocity distribution function (VDF), where the distribution magnitude is small but the associated ion energy is large. The corresponding ion trapping needs to be resolved with a large number of super-particles per cell. Moreover, these instabilities are transient and preclude time averaging as a viable technique for statistical convergence. Such statistical noise is eliminated in the grid-based direct kinetic method, where the distribution function is directly simulated in a discretised phase space without the use of super-particles~\citep[and references therein]{filbet2003comparison, kolobov2007unified, thomas2012review, dimarco2014numerical, hara2018test, palmroth2018vlasov}, keeping in mind that an accurate solution necessitates sufficient resolution in both the physical and velocity space dimensions, the latter particularly to resolve particle trapping~\citep[e.g.,][]{schamel2000hole,dieckmann2004connecting}. We leverage this superiority of the direct kinetic method to gain more precise insights into instability development and transition to turbulence. The direct kinetic method has been shown to yield more accurate linear instability growth rates than the particle-in-cell method due to the statistical noise inherent in the latter, which is seen to vary with the number of particles per cell and the wavenumber of interest~\citep{dieckmann2004simulating,tavassoli2021role}. We verify these growth rates in \S~\ref{sec:stages} and also demonstrate the utility of the large dynamic range of direct kinetic solvers in resolving turbulence structures in \S~\ref{sec:vdf}.
\setcitestyle{notesep={, }}

\subsection{Direct kinetic solver}

The direct kinetic solver employed here was originally developed at the University of Michigan with verification and validation against canonical and complex plasma problems, such as waves, electron-emitting sheaths, and Hall thruster discharges~\citep{hara2012one, hara2014mode, hara2015quantitative, hara2017kinetic, hara2018test, raisanen2019two, vazsonyi2020non}. In contrast to state-of-the-art particle-in-cell solvers, direct kinetic solvers eliminate statistical noise and are suitable for precise investigations of instability growth and turbulence inception as discussed above. For collisionless, unmagnetised plasmas, the solver computes the time evolution of the probability density function, $\tilde{f}_*$, for some particle type~$*=i,e$ (ions, electrons) according to the Vlasov equation
\begin{equation}
\f{\partial \tilde{f}_* \left(\boldsymbol{\tilde{x}},\boldsymbol{\tilde{v}};\tilde{t}\right)}{\partial \tilde{t}} + \boldsymbol{\tilde{v}}\cdot\nabla_{\boldsymbol{\tilde{x}}} \tilde{f}_*\left(\boldsymbol{\tilde{x}},\boldsymbol{\tilde{v}};\tilde{t}\right) + \f{\tilde{q}_* \boldsymbol{\tilde{E}}}{\tilde{m}_*}\cdot\nabla_{\boldsymbol{\tilde{v}}} \tilde{f}_*\left(\boldsymbol{\tilde{x}},\boldsymbol{\tilde{v}};\tilde{t}\right) = 0,
\end{equation}
where $\boldsymbol{\tilde{x}}$, $\boldsymbol{\tilde{v}}$, and $\boldsymbol{\tilde{E}}$ respectively denote the position, velocity, and electric field vectors, $\tilde{q}_*$ denotes the charge of the simulated particle type, and $\tilde{t}$ denotes the time. This is consistent with the conservation of charge in phase space. The computational domain is discretised in $\boldsymbol{\tilde{x}}$--$\boldsymbol{\tilde{v}}$ space with a parallelised second-order finite-volume method, which is described by~\citet{chan2022enabling,chan2022grid}. Assuming small induced magnetic fields and their rates of change, we adopt the electrostatic approximation, where Gauss's law is expressed using $\boldsymbol{\tilde{E}}=-\nabla_{\boldsymbol{\tilde{x}}}\tilde{\phi}$ as a Poisson equation for the electric potential $\tilde{\phi}$ of the form
\begin{equation}
\nabla_{\boldsymbol{\tilde{x}}}^2 \tilde{\phi} = -\f{e\left(\tilde{n}_i-\tilde{n}_e\right)}{\varepsilon_0},
\end{equation}
where $\varepsilon_0$ and $e$ respectively denote the vacuum permittivity and elementary charge, and $\tilde{n}_i$ and $\tilde{n}_e$ respectively denote the ion and electron number densities
\begin{equation}
\tilde{n}_i\left(\boldsymbol{\tilde{x}};\tilde{t}\right) = \int_{\boldsymbol{\tilde{v}}} \tilde{f}_i\left(\boldsymbol{\tilde{x}},\boldsymbol{\tilde{v}'};\tilde{t}\right) \, \diff \boldsymbol{\tilde{v}'}; \qquad 
\tilde{n}_e\left(\boldsymbol{\tilde{x}};\tilde{t}\right) = \int_{\boldsymbol{\tilde{v}}} \tilde{f}_e\left(\boldsymbol{\tilde{x}},\boldsymbol{\tilde{v}'};\tilde{t}\right) \, \diff \boldsymbol{\tilde{v}'}.
\end{equation}
Periodic and no-flux boundary conditions are employed for $\boldsymbol{\tilde{x}}$ and $\boldsymbol{\tilde{v}}$, respectively.

\subsection{Problem set-up and linear stability analysis}\label{sec:linear}

We consider 1D1V and 2D2V current-carrying instabilities, where D and V denote the configuration and velocity spaces, respectively. The corresponding simulations are respectively two- and four-dimensional. Since the initial ion-acoustic wave excites long-wavelength modes, we choose domain lengths sufficiently larger than the Debye length, $\tilde{\lambda}_\text{D} = \sqrt{\left(\varepsilon_0 k_\text{B} \tilde{T}_e\right)/\left(\tilde{n}_e e^2\right)}$. The initial species temperatures are $\tilde{T}_e = 2\text{ eV}$ and $\tilde{T}_i = 0.2\text{ eV}$ in consistency with the temperature ratio described in \S~\ref{sec:intro}. In addition, all species are initialised with Maxwellian velocity distributions. The electrons have a initial net axial drift described by the initial electron Mach number $M_e = \tilde{U}_e/\tilde{c}_e$, which corresponds to a shifted Maxwellian initial condition. The ions have zero initial mean speed, i.e., $M_i = 0$, which corresponds to a stationary Maxwellian initial condition. This configuration is relevant, for example, to hollow cathode plumes, which are biased by direct-current voltages. As such, the initial conditions for this study assume a stable current. This is to be distinguished from counterstreaming studies~\citep[e.g.,][]{forslund1970numerical,bret2008ions,chapman2021nonlinear}. Hereinafter, we define $x$ as the axial direction and $y$ as the transverse direction in 2D. Correspondingly, $v_x$ and $v_y$ denote the axial and transverse velocity dimensions, respectively.

Linear growth rates for the current-carrying instability can be analytically predicted via solution of the following linear dispersion relation, which is derived for one- and multi-dimensional electrostatic waves travelling at an arbitrary angle $\theta$ to the axial direction across the background ion and electron populations defined above (stationary and shifted Maxwellians, respectively) with initial species Mach numbers $M_*$ as follows
\begin{equation}
1 + \sum_* \f{\tilde{\omega}_*^2}{\tilde{c}_*^2\tilde{k}^2}\left[1 + \left(\f{\tilde{\omega}}{\sqrt{2}\tilde{c}_* \tilde{k}} - \f{  M_*\cos\theta}{\sqrt{2}}\right) Z\left(\f{\tilde{\omega}}{\sqrt{2}\tilde{c}_* \tilde{k}}  - \f{  M_*\cos\theta}{\sqrt{2}}\right)  \right] = 0,\label{eqn:disp}
\end{equation}
where $\tilde{k} = \left|\boldsymbol{\tilde{k}}\right|$ and $\tilde{\omega}$ are, respectively, the modal wavenumber magnitude and angular frequency of the electrostatic wave in question, $\tilde{\omega}_* = \sqrt{\left(\tilde{n}_* e^2\right)/\left(\tilde{m}_* \varepsilon_0\right)}$ is the plasma frequency, $\tilde{c}_* = \sqrt{k_\text{B} \tilde{T}_*/\tilde{m}_*}$ is the species thermal speed, and $Z$ is the plasma dispersion function corresponding to a Maxwellian velocity distribution
\begin{equation}
Z(\xi) = \f{1}{\sqrt{\upi}}\int_{-\infty}^{\infty} \f{e^{-z^2}}{z - \xi} \, \diff z = \ii \sqrt{\upi} e^{-\xi^2} \text{erfc}(-\ii \xi); \qquad \f{\diff Z(\xi)}{\diff \xi} = -2\left[1 + \xi Z(\xi)\right].
\end{equation}

Unless otherwise stated, lengths, velocities, and time are hereinafter respectively non-dimensionalised by $\tilde{\lambda}_\text{D}$, $\tilde{c}_*$, and the inverse electron frequency $1/\tilde{\omega}_e = \tilde{\lambda}_\text{D}/\tilde{c}_e$. Specifically, $k = \tilde{k}\tilde{\lambda}_\text{D}$, $\boldsymbol{x} = \boldsymbol{\tilde{x}}/\tilde{\lambda}_\text{D}$, $f_* = \tilde{f}_*\tilde{c}_*$, $\boldsymbol{v} = \boldsymbol{\tilde{v}}/\tilde{c}_*$, and $t = \tilde{\omega}_e\tilde{t}$. Note that the characteristic ion time scale, $1/\tilde{\omega}_i$, and correspondingly the ion-acoustic period, exceed the characteristic electron time scale, $1/\tilde{\omega}_e$, by a factor of $\sqrt{\tilde{m}_i/\tilde{m}_e} \approx 40$.

In this work, we consider the evolution of the following energy modes (all listed in their dimensional and volume-integrated forms): the electrostatic potential energy $\int \f{1}{2} \varepsilon_0 |\mathbf{\tilde{E}}|^2 \, \diff \tilde{V}$, the bulk kinetic energy $\int \f{1}{2} \tilde{n}_* \tilde{m}_* \tilde{U}_*^2 \, \diff \tilde{V}$, and the random kinetic energy $\int \tilde{n}_* k_\text{B} \tilde{T}_* \, \diff \tilde{V}$. In particular, we consider the kinetic energies in detail in \S~\ref{sec:1D}.

\section{2D current-carrying instability}\label{sec:2D}

Building on the preliminary work of~\citet{vazsonyi2021deterministic}, the 2D2V instability is simulated with domain extents $x,y \in [0, L = 80], \, v_{x,i}, v_{y,i} \in [-45, 45], \, v_{x,e}, v_{y,e} \in [-8.5, 8.5]$ and resolutions $\Delta x = \Delta y = 0.4, \, \Delta v_{x,i} = \Delta v_{y,i} = 0.7, \, \Delta v_{x,e} = \Delta v_{y,e} = 0.1$ for the spatial, ion velocity, and electron velocity dimensions, respectively. These are in line with the grid-point recommendations of \citet{chan2022grid} for adequate resolution of macroscopic quantities and gradients. The corresponding number of grid points are $N_x = N_y = 200, \, N_{v_{x,i}} = N_{v_{y,i}} = 128, \, N_{v_{x,e}} = N_{v_{y,e}} = 192$. The simulations were performed on 96 Intel Xeon Gold CPUs over a total wall-clock duration of about 4 months. The baseline simulation is converged with respect to the spatial grid and domain extent (not shown here), as well as the velocity grid (see appendix~\ref{sec:gridconv}), for the macroscopic quantities of interest. For the baseline simulation, $\Delta t = 0.06$ and $M_e=2.3$ using the initial and boundary conditions detailed in \S~\ref{sec:setup}. Hereinafter, electric potentials and field strengths are normalised by the thermal potential, $\tilde{\phi}_\text{th} = k_\text{B}\tilde{T}_e/e$, and the corresponding characteristic field strength, $\tilde{\phi}_\text{th}/\tilde{\lambda}_\text{D}$, respectively. Specifically, $\phi = \tilde{\phi}/\tilde{\phi}_\text{th}$ and $\boldsymbol{E} = \boldsymbol{\tilde{E}}\tilde{\lambda}_\text{D}/\tilde{\phi}_\text{th}$.

\subsection{Stages of instability growth: electric potential and energy spectra}\label{sec:stages}

\begin{figure}
\begin{center}
    	\includegraphics[trim = 0 150 0 140, width = \textwidth]{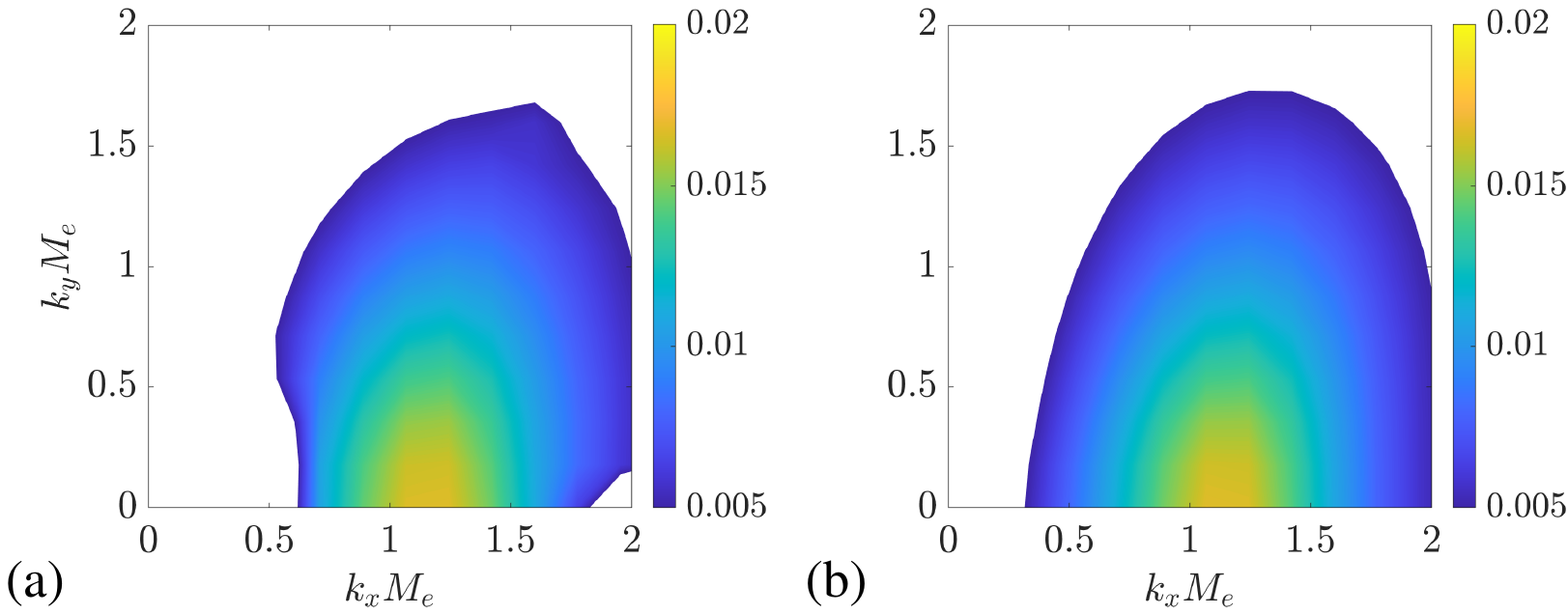}
      	\caption{Numerical growth rates of spectral modes $\{k_x M_e,k_y M_e\}$ for the first quadrant of the potential spectrum $E_{\phi\phi}$, which is obtained via a Fourier transform of $\phi$ in physical space into its Fourier coefficients $\hat{\phi}$, and then computation of $\hat{\phi}\hat{\phi}^*$, which is the modal coefficient multiplied by its complex conjugate. The growth rates are plotted for $t=[3.4\times10^1,2.7\times10^2]$ in panel (a) and are obtained via linear regression of the spectral magnitudes in time. The maximum growth rate predicted by the dispersion relation in \eqref{eqn:disp} is 0.017, and the analytical growth rates corresponding to other modes of interest are obtained through numerical solution of the dispersion relation in \eqref{eqn:disp} and plotted in panel (b). Growth rates less than $5\times10^{-3}$ are excluded to remove cases where oscillations confound the regression. Hereinafter, potentials and wavenumbers are normalised by $\tilde{\phi}_{\text{th}}$ and $1/\tilde{\lambda}_\text{D}$, respectively, and times by $1/\tilde{\omega}_e$.\label{fig:growth2Dlin}}
\end{center}
\end{figure}

\begin{figure}
\begin{center}
    	\includegraphics[trim = 0 150 0 140, width = \textwidth]{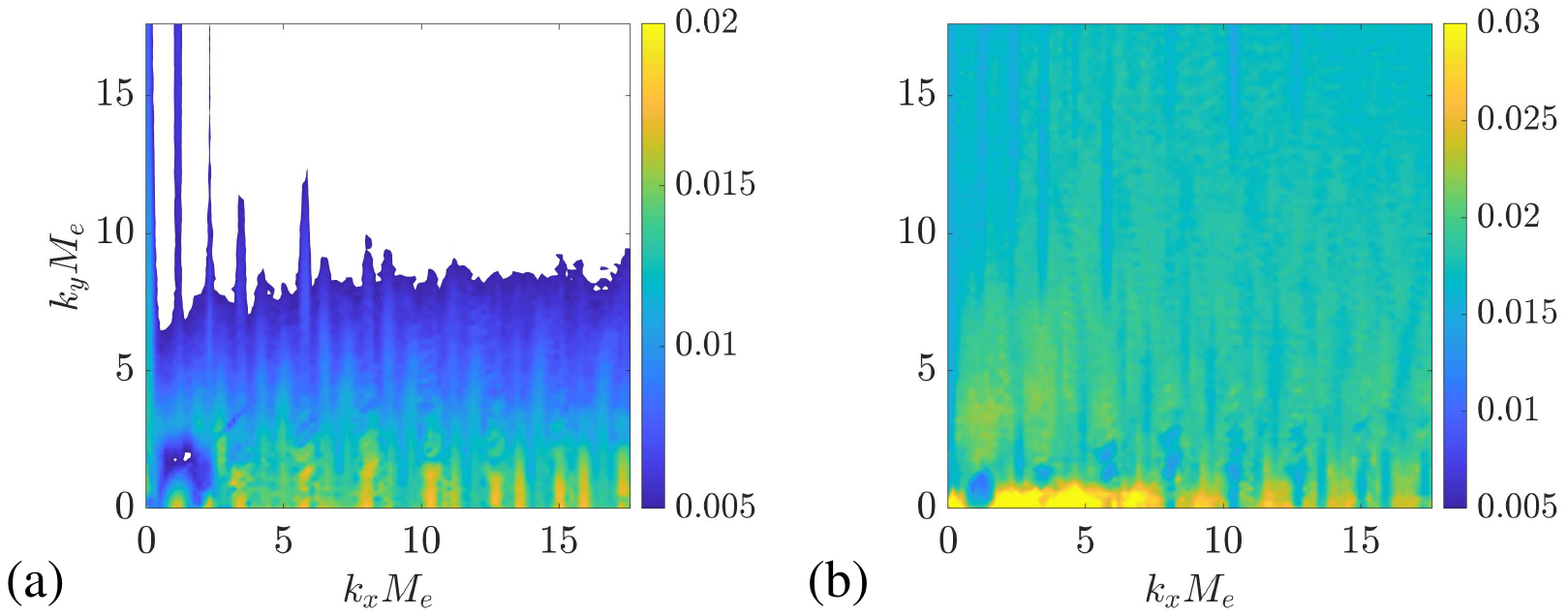}
      	\caption{The same growth rates plotted in figure~\ref{fig:growth2Dlin} for (a) $t=[4.6\times10^2,1.1\times10^3]$ and (b) $t=[1.1\times10^3,1.6\times10^3]$. Note the different colourbar maximum in panel (b). The extent of plotted wavenumbers approaches $k_x, k_y \approx 2\pi$, which corresponds to the Debye length itself. \label{fig:growth2Dnonlin}}
\end{center}
\end{figure}

\begin{figure}
\begin{center}
    	\includegraphics[trim = 0 120 0 110, width = \textwidth]{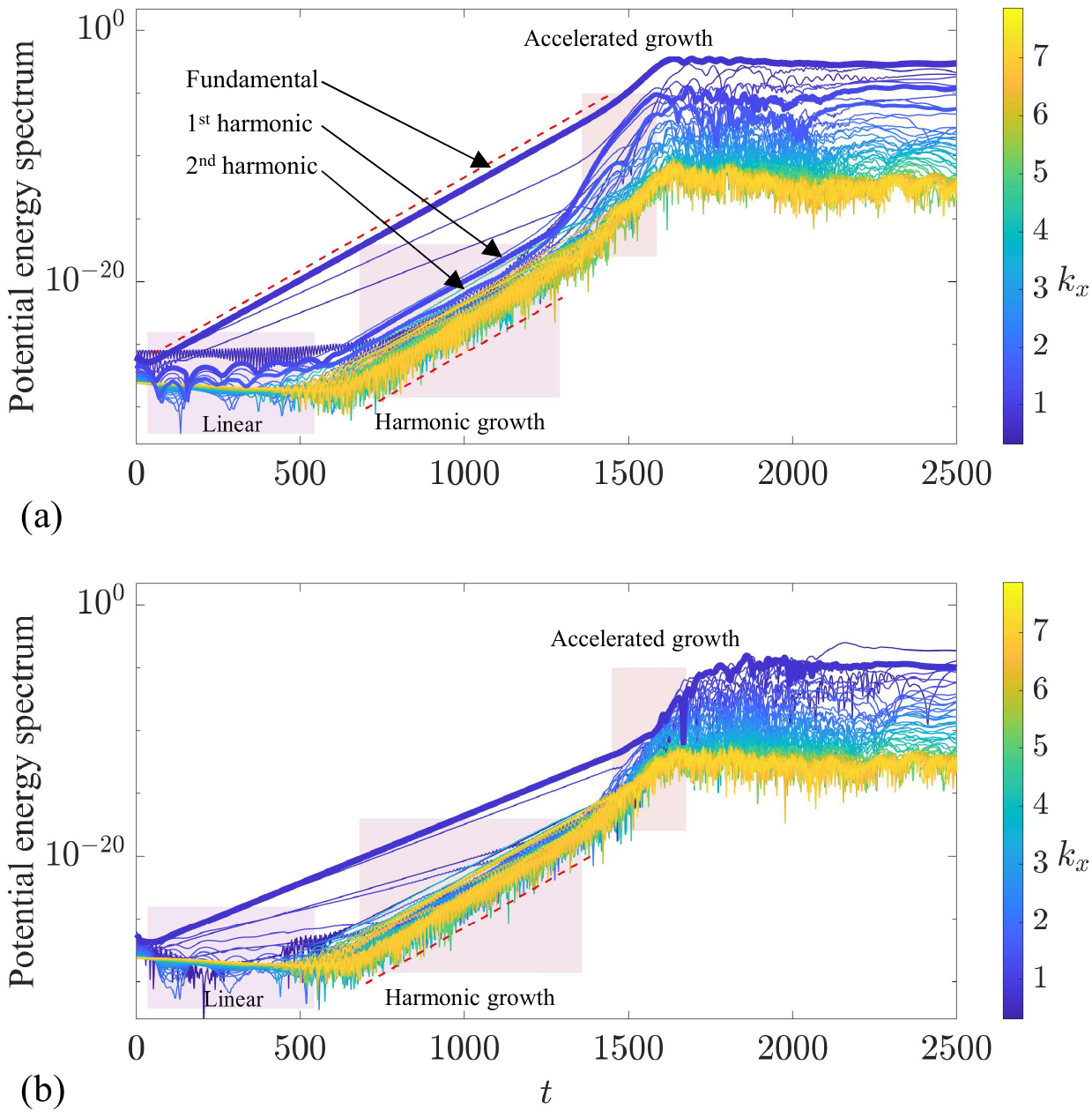}
      	\caption{Time evolution of the electrostatic potential energy spectrum, which is obtained via a Fourier transform of the axial and transverse field strengths in physical space, $E_x$ and $E_y$, into their Fourier coefficients $\hat{E}_x$ and $\hat{E}_y$, and then computation of the quantity $\hat{E}_x\hat{E}_x^* + \hat{E}_y\hat{E}_y^*$. The spectral coefficients are plotted for (a) $k_y = 0$ and (b) $k_y = 0.3$. The sloped dashed lines denote the maximum growth rate from \eqref{eqn:disp}. Every second mode is plotted $(\Delta k_\text{plot} = 2\Delta k = 2 k_\text{min} = 4\pi/80 \approx 0.16)$ and the curves are coloured from blue to yellow (dark to light in greyscale) in increasing $k$ $(k_\text{max} = N_x k_\text{min}/2 \approx 7.9)$. The curves corresponding to the fundamental $x$ wavenumber are bolded in panels (a) and (b), while those corresponding to the first two harmonics are also bolded in panel (a). The shaded regions denoting the instability stage are intended only as visual guides, as the linear, harmonic growth, and accelerated growth stages differ for each mode. Hereinafter, field strengths are normalised by $\tilde{\phi}_{\text{th}}/\tilde{\lambda}_\text{D}$.\label{fig:kx2D}}
\end{center}
\end{figure}

Figures~\ref{fig:growth2Dlin} and \ref{fig:growth2Dnonlin} plot the modal potential growth rates over three time intervals for the baseline simulation introduced above but with twice the velocity resolution. These are based on the potential spectrum $E_{\phi\phi} = \hat{\phi}\hat{\phi}^*$, where the hat notation denotes a Fourier transform and the superscripted asterisk denotes the complex conjugate operation. In the linear stage depicted in figure~\ref{fig:growth2Dlin}, the simulated growth rates in panel (a) agree with the theoretically predicted rates in panel (b), which were obtained through numerical solution of the dispersion relation in \eqref{eqn:disp} using standard iterative techniques for non-linear equations, as well as those reported by~\citet{amano2009nonlinear}. The maximum growth rate occurs at $k_x M_e \approx 1.2$ and matches the corresponding 1D growth rate, since the 2D dispersion relation in \eqref{eqn:disp} is equivalent to its 1D counterpart when $\theta = 0$. By symmetry of the dispersion relation, the growth rates are reflectionally symmetric about the $k_x$ axis, i.e., the wave solution is identical for positive and negative $k_y$ (not shown here). Note that the analysis method adopted here effectively sums the contributions of positive and negative $k_x$ by virtue of the complex conjugate product in the definition of $E_{\phi\phi}$. Conversely, the linear instability is a one-way instability where positive $k_x$ are most unstable. As $k_x$ increases, the growth rates decrease for axial $x$ harmonics of the fastest-growing fundamental $(k_x \neq 0, k_y = 0)$ along the $k_x$ axis, as well as their neighbouring modes. The subsequent stages of the instability development process are depicted in figure~\ref{fig:growth2Dnonlin}. Growth of these harmonics at a comparable rate to the fastest-growing fundamental is first observed in figure~\ref{fig:growth2Dnonlin}(a). At later times, accelerated growth in the forms of quasi-linear harmonic growth~\citep[see][and references therein]{rajawat2017particle} and a strong non-linear fill-in process at intermediate wavenumbers via a catch-up mechanism is then observed in figure~\ref{fig:growth2Dnonlin}(b). Here, primarily non-harmonic wavenumbers experience accelerated growth beyond the linear growth rate of the fastest-growing fundamental mode. Eventually, no further modal growth occurs on average and the potential reaches a saturated turbulence state.

\begin{figure}
\begin{center}
    	\includegraphics[trim = 0 85 0 80, width = \textwidth]{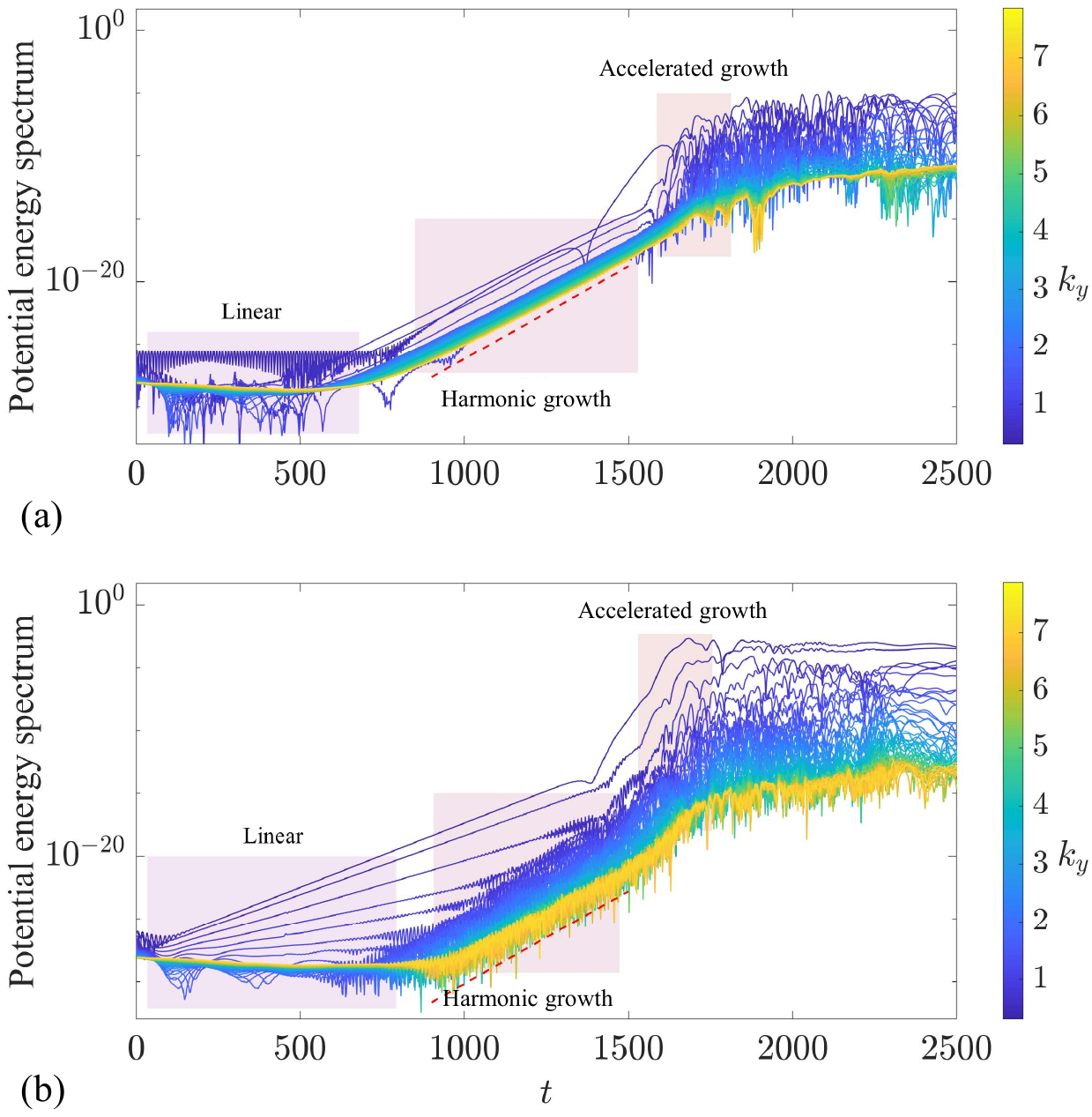}
      	\caption{Time evolution of the electrostatic potential energy spectrum for (a) $k_x = 0$ and (b) $k_x = 0.3$. For a description of the dashed lines and colours, refer to the caption of figure~\ref{fig:kx2D}. \label{fig:ky2D}}
\end{center}
\end{figure}

To further illustrate the transition between the linear, harmonic growth, and accelerated growth regimes of the instability growth, figures~\ref{fig:kx2D} and \ref{fig:ky2D} plot the electrostatic potential energy modal coefficients, $\hat{E}_x\hat{E}_x^* + \hat{E}_y\hat{E}_y^*$, as functions of time spanning the time intervals discussed in figures~\ref{fig:growth2Dlin} and \ref{fig:growth2Dnonlin}, as well as beyond. Here, $E_x$ and $E_y$ respectively denote the axial and transverse electric field strengths. In each panel of figure~\ref{fig:kx2D}, $k_y$ is held fixed while $k_x$ is varied, i.e., the growth of waves of different axial wavenumbers is directly compared. In each panel of figure~\ref{fig:ky2D}, $k_x$ is held fixed while $k_y$ is varied, i.e., the growth of waves of different transverse wavenumbers is directly compared. 

Modes first grow linearly at their analytically predicted modal growth rates, with high-$k$ modes exhibiting slow growth and even decay in agreement with linear stability theory and in contrast to their faster-growing low-$k$ counterparts, as marked by the left shaded region in each panel. Particularly, the growth rate of the fastest-growing mode, as indicated by the darkest bolded curve in figure~\ref{fig:kx2D}(a), matches the theoretical maximum growth rate, as indicated by the red dashed line in the same panel. The curve corresponding to the same axial wavenumber is also bolded in figure~\ref{fig:kx2D}(b), corroborating the observation in figure~\ref{fig:growth2Dlin} that the axial wavenumber of the fastest-growing fundamental is insensitive to the transverse wavenumber. In further corroboration with the growth rates of figure~\ref{fig:growth2Dlin}, modes of various transverse wavenumbers exhibit larger linear growth at $k_x=0.3$ than at $k_x=0$, as shown in figure~\ref{fig:ky2D}. The duration of the linear growth phase depends on the wavenumber magnitude. This is approximately visualised by the different time extents of the left shaded regions in figures~\ref{fig:kx2D} and \ref{fig:ky2D}, although the shaded regions are only intended as visual guides since the phase duration differs for each individual mode. 

Subsequently, harmonics of the fastest-growing fundamental start to grow at a comparable (but slightly slower) rate to the fundamental, as marked by the middle shaded region in each panel. This process begins with the axial $x$ harmonics [$k_x \neq 0, k_y = 0$, figure~\ref{fig:kx2D}(a)]: as also evidenced by figure~\ref{fig:growth2Dnonlin}(a), purely axial modes whose $x$ wavenumbers are approximately integer-valued multiples of the most rapidly growing fundamental lead the harmonic growth process. Similar dynamics occur for weakly oblique $x$ harmonics [figure~\ref{fig:kx2D}(b)], where the $x$ wavenumbers are identically almost integer-valued multiples of the fastest-growing fundamental while the $y$ wavenumbers are slightly non-zero. Figure~\ref{fig:ky2D} indicates that this is then followed by the purely transverse $(k_x = 0, k_y \neq 0)$ and eventually strongly oblique $(k_x, k_y \neq 0\text{ harmonics where both wavenumber components have large magnitudes})$ resonant modes. The observed delay in this onset is about 100--200 dimensionless times in figure~\ref{fig:ky2D}(a) and 300--500 dimensionless times in figure~\ref{fig:ky2D}(b). Note that the $k_x=0$ purely transverse modes in figure~\ref{fig:ky2D}(a), which experienced limited growth in the linear phase, are eventually excited to significant amplitudes in the harmonic growth phase. 

The final accelerated growth phase is indicated approximately by the right shaded regions in figures~\ref{fig:kx2D} and \ref{fig:ky2D}. Here, the remaining non-harmonic modes, which include both axis-parallel and oblique wavevector orientations, experience accelerated super-linear growth, i.e., beyond that predicted for any mode by linear theory. This abrupt growth fills in the intermediate wavenumbers to form a saturated and persistent broadband spectrum. While the non-harmonic modes grow significantly in this phase to catch up to the harmonic and fundamental wave amplitudes in the broadband spectrum, the growth is much less pronounced for these latter wave categories. The time interval between the onset of harmonic growth and the development of saturated turbulence in the electric potential and field is about $O(10^3)$ dimensionless times, or $O(10)$ ion-acoustic periods.

\begin{figure}
\begin{center}
    	\includegraphics[trim = 0 0 0 0, width = 0.65\textwidth]{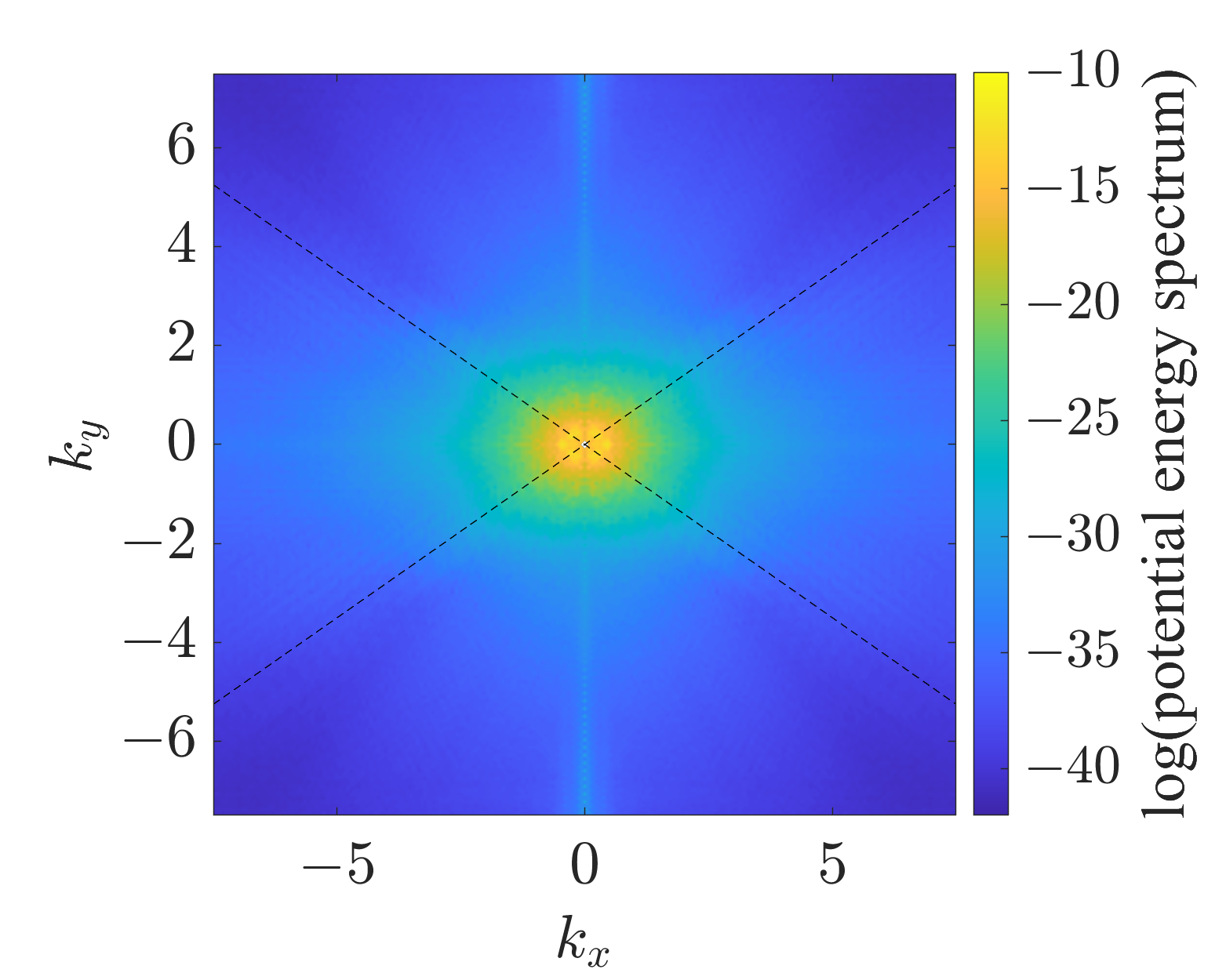}
      	\caption{Electrostatic potential energy spectrum in the non-linear saturation phase averaged over the time interval $t \in [2.9 \times10^3,3.7 \times10^3]$. The dashed lines mark 35\degree~angles from the horizontal. \label{fig:satcoeff}}
\end{center}
\end{figure}

The aforementioned growth of harmonics (locking) and subsequent fill-in (non-linear regime) are features of turbulence also seen in, e.g., hydrodynamic simulations of turbulent boundary layers, where energy cascades from long-wavelength modes to their short-wavelength harmonics through quasi-linear interactions, while non-linear triadic and polyadic interactions induce eventual fill-in of the intermediate-wavenumber spectral content~\citep[e.g.,][]{sayadi2011direct}. Parallel analysis of an ensemble of comparable 1D simulations (not shown here) suggests that the growth rates and order of growth of modes are insensitive to the details of the initial system noise keeping its magnitude constrained. As with hydrodynamic turbulence, this suggests that the system has limited memory of the initial conditions, as self-similarity eventually percolates across the system scales to achieve some form of statistical equilibrium. Note that longer modes ($k \leq 1$) mostly lead and exceed shorter modes ($k \geq 1$) in the modal growth process illustrated in figure~\ref{fig:kx2D} and particularly figure~\ref{fig:ky2D}, possibly casting in doubt the existence of an inverse energy cascade [see, e.g., the discussion of causality in modal development by~\citet{lozanoduran2022information}]. In other words, strong parallels with hydrodynamic turbulence are exhibited, where the dominant dynamics are associated with a forward energy cascade predominantly carrying energy from large to small scales. As such, electrostatic plasma turbulence of a sufficiently large separation of scales should also reasonably satisfy local isotropy, where small- and intermediate-scale statistics are mostly independent of large-scale forcing details, and are thus weak functions of spatial direction in small spatial neighbourhoods. This is largely supported by the time-averaged electrostatic potential energy spectrum in its saturated turbulence state, as depicted in figure~\ref{fig:satcoeff}. The wave energy is maximum at the lower-$k$ modes (i.e., long-wavelength modes) and decays toward high-$k$ modes. There exists a minor tendency for waves to travel at an angle of 35\degree~to the axial direction due to ion-induced wave scattering. When the electric field is oriented in this direction, its axial component is slightly larger than its transverse component, which is in corroboration with the initial instability being driven in the axial direction. In addition, this tendency is consistent with the analytical prediction in the review by~\citet{bychenkov1988ion}, which discusses the role of ion-induced scattering of ion-acoustic waves on turbulent fluctuations in more detail. Apart from this minor asymmetry, the energy spectrum exhibits isotropy at intermediate scales on the whole. Such local isotropy is consistent with the principles of scale separation and self similarity underlying a turbulent field with largely uni-directional inter-scale transport from large to small scales, as introduced in the seminal works of~\citet{richardson1922weather}, \citet{kolmogorov1941local}, and \citet{onsager1945distribution} in the context of the classical energy cascade in hydrodynamic turbulence [for generalisation beyond kinetic energy transport, see, e.g.,~\citet{chan2021turbulent}]. It should be emphasised that the current work considers unmagnetised instabilities where there is no bulk magnetic field introducing further anisotropy into the system. This study focuses on possible connections and similarities between hydrodynamic and electrostatic plasma turbulence as a starting point to invoke parallels in any cascading dynamics, and the effects of magnetic anisotropy are to be a subject of further study.

\subsection{Turbulent saturation: ion velocity and energy distributions}\label{sec:vdf}

\begin{figure}
\begin{center}
    	\includegraphics[trim = 0 60 0 50, width = \textwidth]{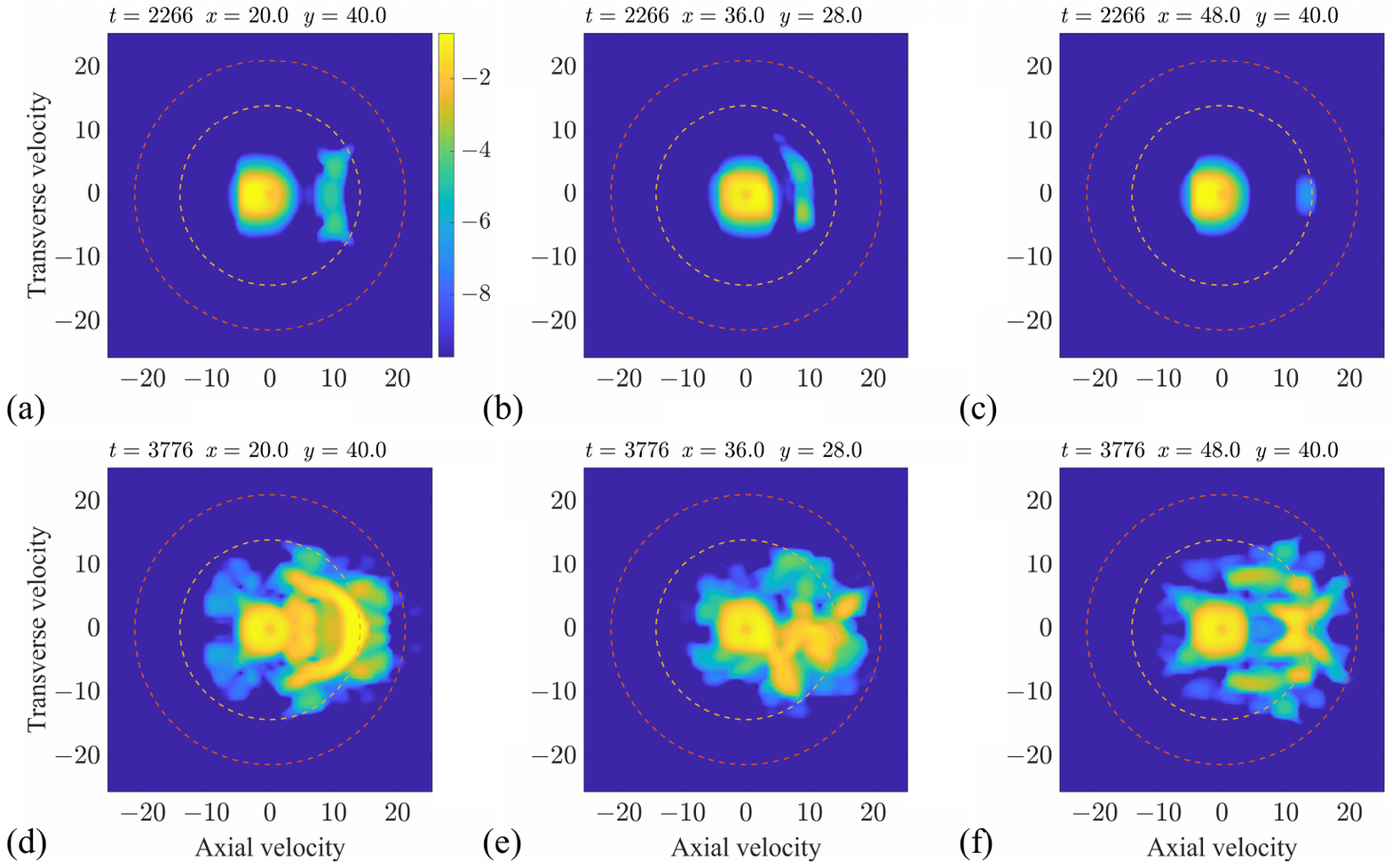}
      	\caption{Local ion velocity distribution functions, weighted by the squared velocity magnitude to highlight the phase-space distribution of energy, at three different locations at (a--c) $t = 2.3\times10^3$ and (d--f) $t = 3.8\times10^3$. The contours are logarithmically spaced with their corresponding exponents labelled in the colourbar and the two concentric circles represent 20 and 45 eV contours. Hereinafter, velocities are normalised by the species thermal speed $\tilde{c}_*$.\label{fig:vdf2Dlocal}}
\end{center}
\end{figure}

\begin{figure}
\begin{center}
    	\includegraphics[trim = 60 0 70 0, width = \textwidth]{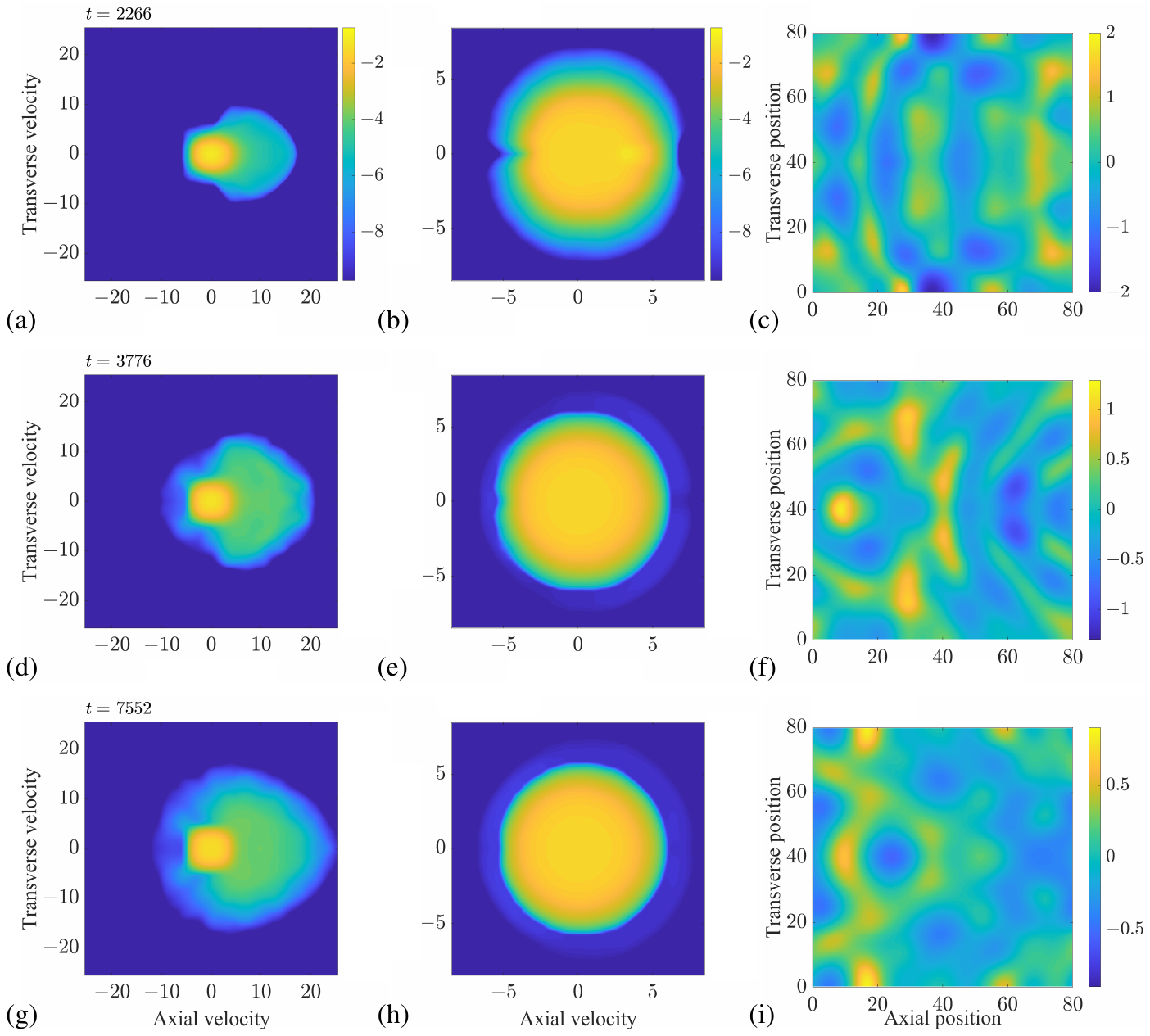}
      	\caption{Time evolution of spatially averaged (a,d,g) ion and (b,e,h) electron velocity distribution functions (VDFs), as well as (c,f,i) the electric potential field. The provided snapshots are (a--c) near the time instant where the potential reaches its saturated turbulence state, as well as about (d--f) 35 and (g--i) 120 ion-acoustic times after. \label{fig:vdf2DtimeM2p3}}
\end{center}
\end{figure}

Figure~\ref{fig:vdf2Dlocal} plots the ion velocity distribution, weighted by the squared velocity magnitude, i.e., $|\boldsymbol{v}|^2 f_i(\boldsymbol{v})$, to highlight the phase-space distribution of energy. The results for the baseline simulation introduced in \S~\ref{sec:2D} are shown at three spatial locations for two time instances soon after the potential reaches its saturated turbulence state as observed in \S~\ref{sec:stages}. Shortly after potential saturation, figures~\ref{fig:vdf2Dlocal}(a)--(c) demonstrate high-energy ion formation in the direction of the net electron drift at $t=2.3\times10^3$, which is past the accelerated growth stage identified in figures~\ref{fig:kx2D} and \ref{fig:ky2D}, with similarities to the 1D instability~\citep{hara2019ion}. The energetic ions adopt speeds that exceed the initial ion-acoustic speed, which is approximately $3\tilde{c}_i$ for the chosen $\tilde{T}_e/\tilde{T}_i$, by under an order of magnitude. While these elevated speeds are suggestive of electron heating (see also figure~\ref{fig:vdf2DtimeM2p3}), the ions remain strongly resonant with dominant ion-acoustic modes and exhibit coherence in physical space at these early times when not many ion-acoustic periods have elapsed since potential saturation. Conversely, at $t=3.8\times10^3$, i.e., about 30 ion-acoustic periods later, figures~\ref{fig:vdf2Dlocal}(d)--(f) indicate that more multi-scale ion phase-space structures, i.e., with a broad range of characteristic scales, are observed. High-energy ions are generated in significant amounts, first in the forward ($+x$) direction, and subsequently in the transverse ($\pm y$) and backward ($-x$) directions due to the multi-dimensional plasma wave formation, at equivalent temperatures of between 20 and 50 eV. The velocity distributions now demonstrate greater incoherence in physical space due to accumulated response to the sustained broadband spectrum of ion-acoustic modes. Once again, the corresponding energetic ion speeds exceed the initial ion-acoustic speed and may be associated with concomitant electron heating. Such elevated energies can be of interest to downstream physical and engineering phenomena, such as the onset of hollow cathode sputtering in spacecraft thrusters. However, at the selected initial $M_e$, more backward-streaming ions were observed in the 1D instability~\citep{hara2019ion} and appear to be less abundant in the 2D case.

\begin{figure}
\begin{center}
    	\includegraphics[trim = 110 25 110 0, width = \textwidth]{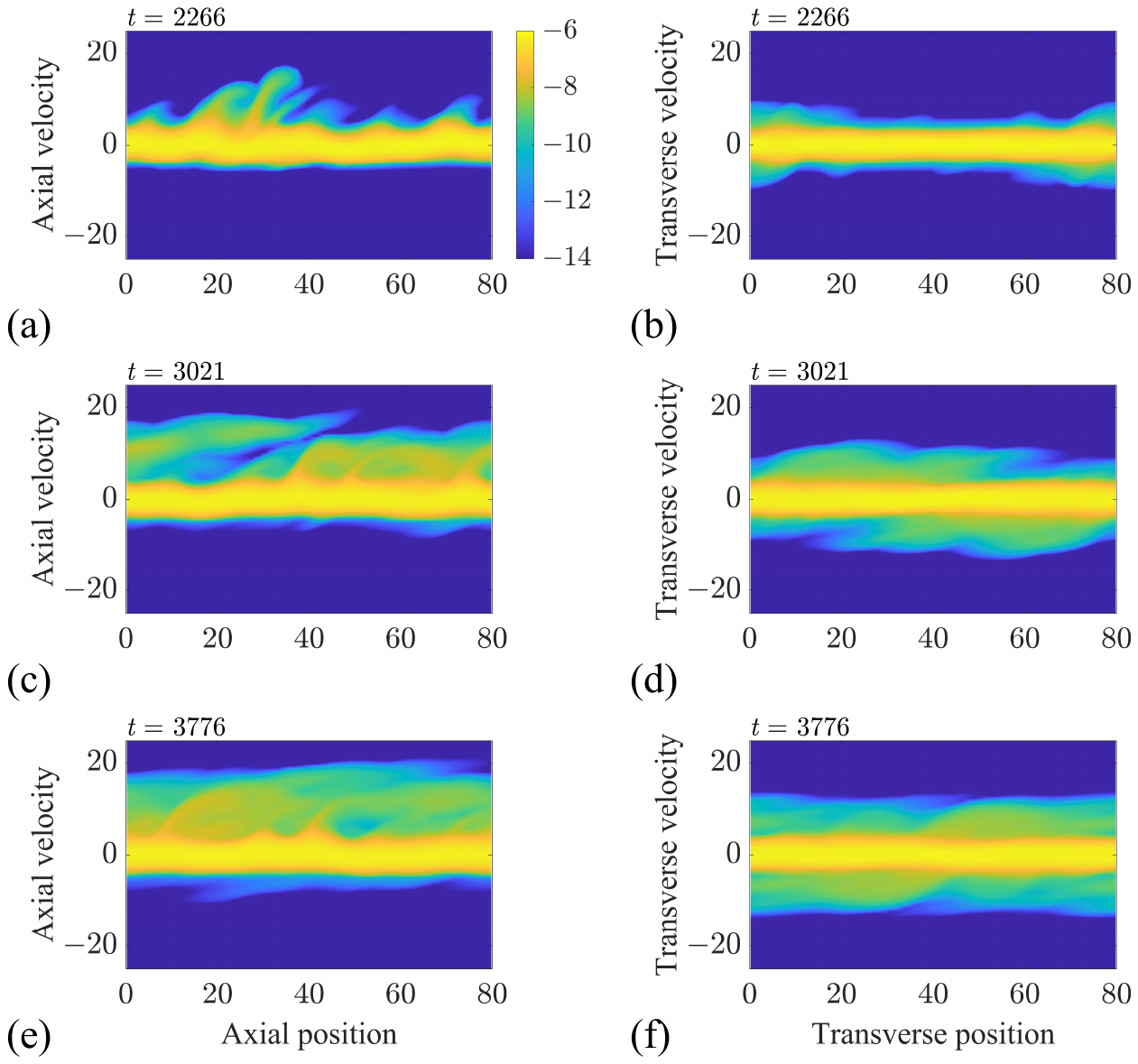}
      	\caption{Ion distribution functions on spatial--velocity axes in the (a,c,e) longitudinal ($x$, $v_x$) and (b,d,f) transverse ($y$, $v_y$) directions, after averaging over the remaining two phase-space axes, at (a,b) $t = 2.3\times10^3$, (c,d) $t = 3.0\times10^3$, and (e,f) $t = 3.8\times10^3$.\label{fig:vdf2Dxvx}}
\end{center}
\end{figure}

In order to depict vortical structures in phase space more clearly, figure~\ref{fig:vdf2Dxvx} plots the time evolution of the ion distribution function on spatial--velocity axes in the longitudinal $\left[\ov{f}_i^{y,v_y}(x,v_x)\right]$ and transverse $\left[\ov{f}_i^{x,v_x}(y,v_y)\right]$ directions in each row of panels, respectively, with averaging $\left(\ov{\cdot}\right)$ performed over the remaining two phase-space axes (in superscript in the compact notation). The snapshots are extracted from the non-linear saturation phase across a duration of about 30 ion-acoustic periods. Figure~\ref{fig:vdf2Dxvx}(a) suggests that plasma waves initially propagate with net motion in the forward axial direction in agreement with 1D theory and observations~\citep{hara2019ion}. Conversely, figure~\ref{fig:vdf2Dxvx}(b) shows that the wave energy along the transverse direction is initially weak, implying insignificant transverse ion trapping. As the saturated plasma waves evolve, the increasing emergence of overturning phase-space structures is indicative of gradually intensifying longitudinal particle trapping as observed in figure~\ref{fig:vdf2Dxvx}(c). The ensuing heating broadens the distribution tails in velocity space. At this time, ions travelling along the transverse direction also demonstrate gradually intensifying trapping in both the forward and backward directions, as indicated in figure~\ref{fig:vdf2Dxvx}(d), with comparable maximum energies along both the transverse and longitudinal directions. The observed structures correspond to waves of a travelling nature along the axial direction and of a standing nature along the transverse direction. As shown in figures~\ref{fig:vdf2Dxvx}(e) and (f), the increasing homogeneity in physical space is a signature of particle trajectories spanning an increasingly isotropic phase-space distribution. It should be noted that there remains a strong signature of ions travelling in the forward axial direction even at these late times, as shown in figure~\ref{fig:vdf2Dxvx}(e), while fast backward-moving ions are observable but not in excessive amounts. This is in contrast with corresponding 1D simulations, where a significant population of high-energy ions eventually travels in the direction opposite to the initial net electron drift~\citep{hara2019ion}. The differences between 1D and 2D simulation results will be discussed in more detail in \S~\ref{sec:1D} with reference to exchanges between various energy modes.

As the fidelity of the phase-space distributions can be further improved with increased velocity resolution due to the filamentous nature of the Vlasov equation, we focus on qualitative trends here and defer a more detailed analysis to future work. In line with this approach, figure~\ref{fig:vdf2DtimeM2p3} plots the time evolution of the spatially averaged ion $\left[\ov{f}_i^{x,y}(v_x,v_y)\right]$ and electron $\left[\ov{f}_e^{x,y}(v_x,v_y)\right]$ velocity distributions, as well as the electric potential ($\phi$), all for the baseline simulation as well. Electron trapping and isotropisation occur more rapidly than their corresponding ion processes owing to the faster thermal speed and shorter response time of electrons. Particle trapping occurs at the phase speed of the excited plasma waves, which is on the order of the ion-acoustic speed $\sqrt{k_\text{B}\tilde{T}_e/\tilde{m}_i}$ and becomes respectively $O(10^{-1})$ and $O(10)$ on the non-dimensional electron and ion velocity axes for the subsequently heightened $\tilde{T}_e$ posited in the discussion of figure~\ref{fig:vdf2Dlocal}. The isotropisation of both velocity distributions is centred around these characteristic speeds in the positive axial direction, continuing the particle trapping process observed at early times in figure~\ref{fig:vdf2Dxvx}. However, note that while plasma waves are observed along both the transverse and longitudinal directions in figures~\ref{fig:vdf2DtimeM2p3}(c,f,i), the wave amplitude is dampened at later times due to energy exchange with particles, and the plasma waves become less coherent. 

Shortly after the potential reaches a saturated turbulence state as indicated in figure~\ref{fig:vdf2DtimeM2p3}(c), the ion phase-space distribution remains largely Maxwellian with the nascent generation of high-energy forward-streaming ions in figure~\ref{fig:vdf2DtimeM2p3}(a). The net electron drift can still be observed in figure~\ref{fig:vdf2DtimeM2p3}(b) in corroboration with the nascent axial ion acceleration in figure~\ref{fig:vdf2Dxvx}(a). Then, the isotropisation of the spatially averaged electron phase-space distribution illustrated in figure~\ref{fig:vdf2DtimeM2p3}(e) is accompanied by an increase in transverse-streaming high-energy ions in figure~\ref{fig:vdf2DtimeM2p3}(d) in corroboration with the heightened ion velocities in figure~\ref{fig:vdf2Dxvx}(e). As observed in figures~\ref{fig:vdf2DtimeM2p3}(g,h) and later times (not shown here), this eventually results in the appearance of backward-streaming ions, as well as higher-energy forward-streaming ions, while the phase-space distributions themselves approach a quasi-steady saturated turbulence state. In summary, the development of multi-scale ion phase-space structures and accompanying cascades, if any, occurs \emph{after} the potential reaches a saturated turbulence state and the electron phase-space distribution obeys isotropic statistics in the absence of any magnetic fields. It may then be reasonable to assume that multi-scale ion phase-space transport occurs in the background of a turbulent potential field with statistically isotropic electrons. The generation of the corresponding broadband ion phase-space structures occurs over a duration of approximately $O(10^4)$ dimensionless times, or $O(100)$ ion-acoustic periods---an order of magnitude larger than the time taken for the electric potential and field to reach a saturated turbulence state. 

\begin{figure}
\begin{center}
    	\includegraphics[trim = 0 165 0 145, width = \textwidth]{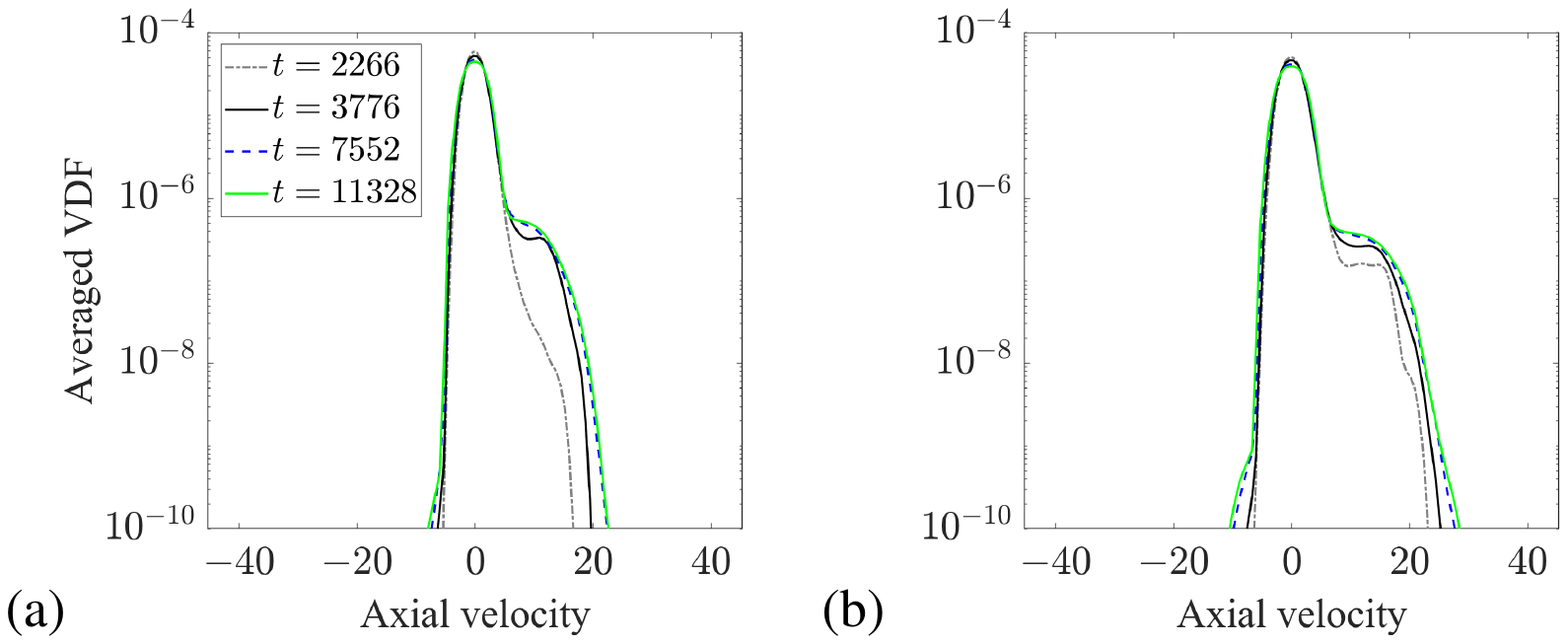}
      	\caption{Time evolution of the axial ion velocity distribution function in the non-linear saturation phase, averaged over physical space and integrated over all transverse velocities, for (a) $M_e = 2.3$ and (b) $M_e = 2.8$. The first three time snapshots correspond to those in figure~\ref{fig:vdf2DtimeM2p3}, while the final time instant is about 210 ion-acoustic times after the onset of potential turbulence in the first snapshot.\label{fig:vdf2Dspaavg}}
\end{center}
\end{figure}

Figure~\ref{fig:vdf2Dspaavg} plots the time evolution of the axial ion velocity distribution function, averaged over physical space and integrated over all transverse velocities, i.e., $\ov{f}_i^{x,y,v_y}(v_x)$, for two $M_e$ values ($M_e = 2.3$ and $2.8$). The ion distributions with a larger initial $M_e$ deviate more rapidly from a Maxwellian and exhibit broader tails, particularly in the backward direction. These tails are evident five to six orders of magnitude lower than the peak distribution values, which necessitate similar variations in the particle-to-cell ratio for adequate resolution by particle solvers and showcase the relative superiority of the direct kinetic solver in this respect~\citep{hara2019overview}. For both $M_e$, the distributions are seen to approach an asymptotic state by the final plotted time instant, indicating statistical saturation in the phase-space distribution itself. An interesting observation is the transient appearance of a forward-streaming distribution plateau at about $t = 4\times10^3$ that steepens at late times. The impermanence of the plateau is likely a consequence of the multi-dimensional nature of the system and can be explained as follows. Strong trapping primarily occurs in the forward-streaming direction at early times, resulting in a discernible distribution plateau similar to that observed in 1D simulations~\citep[see also figure~\ref{fig:FVI1D2D}(a) of this manuscript]{hara2019ion}. A subsequent tendency to statistical isotropy at late times then promotes trapping in more off-axis velocity directions at the expense of axial trapping. Additionally trapped ions can take oblique velocities with axial components smaller than the mean speed corresponding to the original plateau, thereby adding contributions to the left of the plateau and steepening it. The weakening of the plateau also implies continued damping of the plasma waves and is congruent with their gradually decreasing amplitude at late times [see also figure~\ref{fig:vdf2DtimeM2p3}(i) and the time evolution of the electrostatic potential energy in figure~\ref{fig:epe2D} in appendix~\ref{sec:gridconv}]. 

It is interesting to note that the formation of backward-streaming ions is also present in the 2D case but with much smaller amplitude than was observed in its 1D counterpart~\citep{hara2019ion}. In the latter, it was predicted and shown that $M_e \geq 1.3$ results in a transition from the current-carrying ion-acoustic instability (leading to a uni-directional plasma wave) to the Buneman instability (leading to a bi-directional plasma wave in 1D). A key hypothesis of this study was that a similar transition to the Buneman instability would be observed in 2D. However, figure~\ref{fig:vdf2Dspaavg} indicates that the plasma wave amplitude is not large enough to excite enough ions with negative axial velocities. This is in large part due to the nature of energy transfer in multi-dimensional instabilities. 

\section{Comparisons between 1D and 2D instabilities}\label{sec:1D}

In order to relate key conclusions from previous 1D work to their multi-dimensional counterparts, it is instructive to compare several metrics, such as exchanges between different energy modes and macroscopic quantity variations, between the current 2D simulations and their 1D analogues. These metrics enable more insights into the temporal evolution of the ion and electron distribution functions, as well as their corresponding electric potential. Comparable 1D simulations of the same domain extents and resolutions as the baseline 2D simulations introduced in \S~\ref{sec:2D} were performed to this end. We focus on the $M_e = 2.8$ case in this section for brevity.

\subsection{Energy exchange: transfer between different modes}\label{sec:exchange}

We will analyse the transfer of energy between the modes defined in \S~\ref{sec:linear} and reiterated here: the electrostatic potential energy $\int \f{1}{2} \varepsilon_0 |\mathbf{\tilde{E}}|^2 \, \diff \tilde{V}$, the bulk kinetic energy $\int \f{1}{2} \tilde{n}_* \tilde{m}_* \tilde{U}_*^2 \, \diff \tilde{V}$, and the random kinetic energy $\int \tilde{n}_* k_\text{B} \tilde{T}_* \, \diff \tilde{V}$.

\begin{figure}
\begin{center}
    	\includegraphics[trim = 90 10 110 5, width = \textwidth]{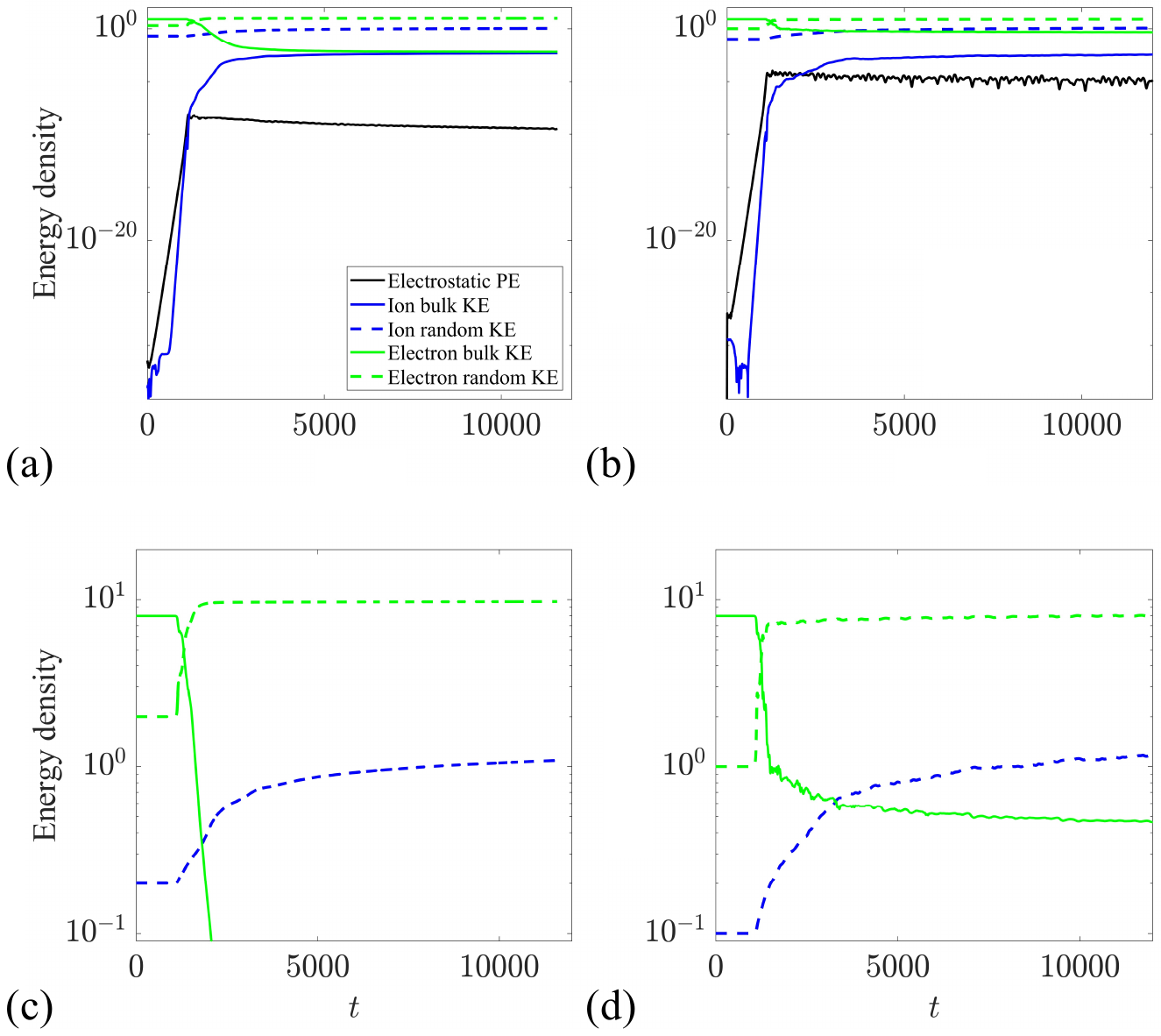}
      	\caption{Time evolution of potential (PE) and various kinetic (KE) energy modes in the (a,c) 2D and (b,d) 1D instabilities for $M_e = 2.8$. The energy densities are plotted on a logarithmic axis with a zoomed-in vertical axis for panels (c,d). Axial and transverse energy densities are summed for the 2D case. All energy densities are computed by dividing the relevant dimensional energies by the domain measure ($L\tilde{\lambda}_\text{D}$ in 1D and $\left[L\tilde{\lambda}_\text{D}\right]^2$ in 2D) and then normalising by the initial electron random kinetic energy density for a single spatial degree of freedom, $\tilde{n}_e k_\text{B} \tilde{T}_e/2$.\label{fig:energetics1D2D}}
\end{center}
\end{figure}

Figure~\ref{fig:energetics1D2D} plots the time evolution of the exchange between different modes of energy for the larger initial electron Mach number considered in \S~\ref{sec:vdf}, in a manner analogous to the energy decomposition considered by~\citet{chan2022enabling}. For both dimensionalities, energy is initially transferred from electron bulk kinetic energy (due to the initial non-zero $M_e$) to the plasma waves, leading to an exponential growth of the electrostatic potential energy. The plasma waves then perturb the ion motion, increasing the ion bulk kinetic energy (particularly through ion trapping at positive axial velocities). Ion and electron heating are simultaneously observed as the plasma waves increase the ion and electron random kinetic energies in tandem. The primary net energy exchange transaction is from electron bulk kinetic energy to electron random kinetic energy with some minor ion heating. 

The ion bulk kinetic energies evolve in a similar fashion in 1D and 2D. However, in 2D, the ion bulk kinetic energy consistently exceeds the electrostatic potential energy following plasma wave saturation, as shown in figure~\ref{fig:energetics1D2D}(a), while it only does so at late times in 1D, as shown in figure~\ref{fig:energetics1D2D}(b). The saturated electrostatic potential energy is much larger in 1D, indicating that the 1D instability results in plasma waves with larger amplitudes that eventually trap ions in both axial directions. The more oscillatory electrostatic potential energy curve in 1D may be attributed to this standing nature of the plasma waves, in contrast to the travelling nature of the axial 2D plasma waves noted earlier in figure~\ref{fig:vdf2Dxvx}. With fewer accessible degrees of freedom in 1D, the ion and electron random kinetic energies reach larger multiples of their initial values. This larger increase in effective ion and electron temperatures in the 1D instability is also driven by stronger electric fields and potential amplitudes as noted above. In the 2D case, the saturated electrostatic potential energy and electron bulk kinetic energy are concurrently lower, as the initial energy in the system is transferred to more degrees of freedom in the ion and electron random kinetic energies, and the magnitude of energy transfer is larger in absolute terms in the 2D instability.

\begin{figure}
\begin{center}
    	\includegraphics[trim = 0 150 0 145, width = \textwidth]{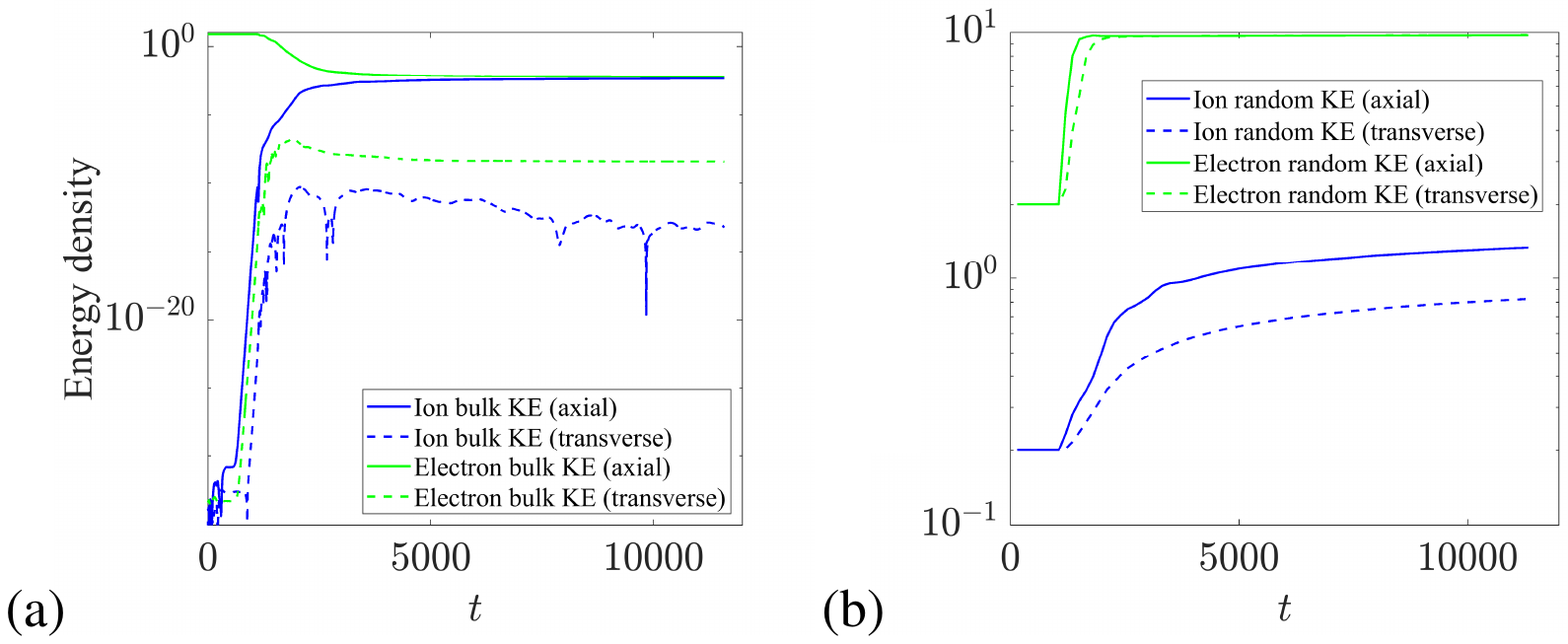}
      	\caption{Time evolution of the axial and transverse kinetic energy (KE) modes for the (a) bulk and (b) random energies in the 2D2V $M_e = 2.8$ case. Note the difference in vertical axis ranges in the two panels. For the energy normalisation, refer to the caption of figure~\ref{fig:energetics1D2D}.\label{fig:energetics2Dsplit}}
\end{center}
\end{figure}

Figure~\ref{fig:energetics2Dsplit} plots the decomposition of the ion and electron kinetic energies into their axial and transverse components for the 2D instability. These may be written dimensionally as
\begin{equation*}
    \int \f{1}{2} \tilde{n}_* \tilde{m}_* \tilde{U}_{*,x\text{ or }y}^2 \, \diff \tilde{V}
\end{equation*}
for the bulk energy components and 
\begin{equation*}
    \int \f{1}{2} \tilde{n}_* k_\text{B} \tilde{T}_{*,x\text{ or }y} \, \diff \tilde{V}
\end{equation*}
for the random energy components. For the ion and electron bulk kinetic energies shown in figure~\ref{fig:energetics2Dsplit}(a), the axial energy dominates the transverse energy throughout the instability growth and saturation. Note from the transverse energy curves that anisotropy appears early in the system in the process of turbulent saturation, where specific modes dominate prior to the non-linear fill-in stage and the potential spectrum is not uniformly broadband. Some isotropisation in species motion then occurs at later times amid the isotropic turbulent potential field to reduce the mean transverse velocities. For the ion and electron random kinetic energies shown in figure~\ref{fig:energetics2Dsplit}(b), axial heating occurs before and more intensely than transverse heating. The two energy components quickly equilibrate in the case of the electrons, marking a rapid return to statistical isotropy. On the other hand, the ions clearly exhibit a reduced tendency to isotropy, which is indicative of lower trapping efficiency away from the forward axial direction.

\subsection{Distribution isotropy and mean drifts}

\begin{figure}
\begin{center}
    	\includegraphics[trim = 0 165 0 145, width = \textwidth]{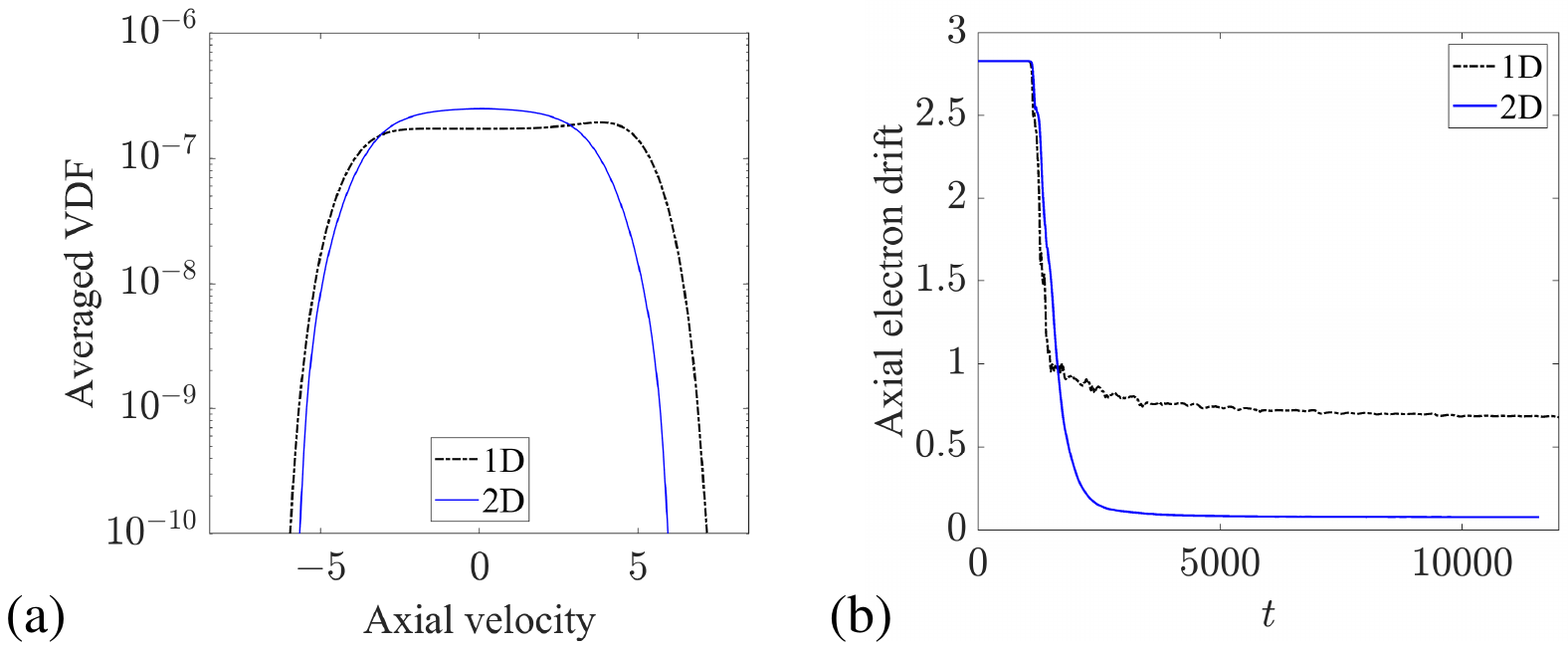}
      	\caption{(a) Comparison of 1D and 2D axial electron velocity distribution functions (VDFs), averaged over all points in space and $t \in [7\times10^3,9\times10^3]$ in the non-linear saturation regime, and integrated over all transverse velocities for the 2D VDF. (b) Comparison of time evolution of 1D and 2D axial electron drifts. Both plots are for $M_e = 2.8$. \label{fig:FVE1D2D}}
\end{center}
\end{figure}

Figure~\ref{fig:FVE1D2D}(a) plots a comparison of the averaged axial electron velocity distribution functions, $\ov{f}_e^{x,y,v_y}(v_x)$, for the 1D and 2D instabilities. The 1D distribution extends to larger speeds as an indication of a larger effective electron temperature, in corroboration with the discussion of figure~\ref{fig:energetics1D2D}. At the late times in the non-linear saturation regime considered here, the 1D distribution retains a peak around in the positive axial direction exhibiting the effects of the initial electron drift. This is absent in the 2D distribution, indicating that the initial electron drift energy is almost entirely transferred to the multi-dimensional plasma waves alongside electron heating. It should be emphasised that there is an additional degree of freedom for the deposition of random kinetic energy in the 2D case that constitutes strong heating in absolute terms. Owing to the multi-dimensional nature of the 2D instability, the phase-space distribution plateau is allowed to extend in multiple dimensions, resulting in the lower net electron axial drift seen in figure~\ref{fig:FVE1D2D}(b).

\begin{figure}
\begin{center}
    	\includegraphics[trim = 0 165 0 145, width = \textwidth]{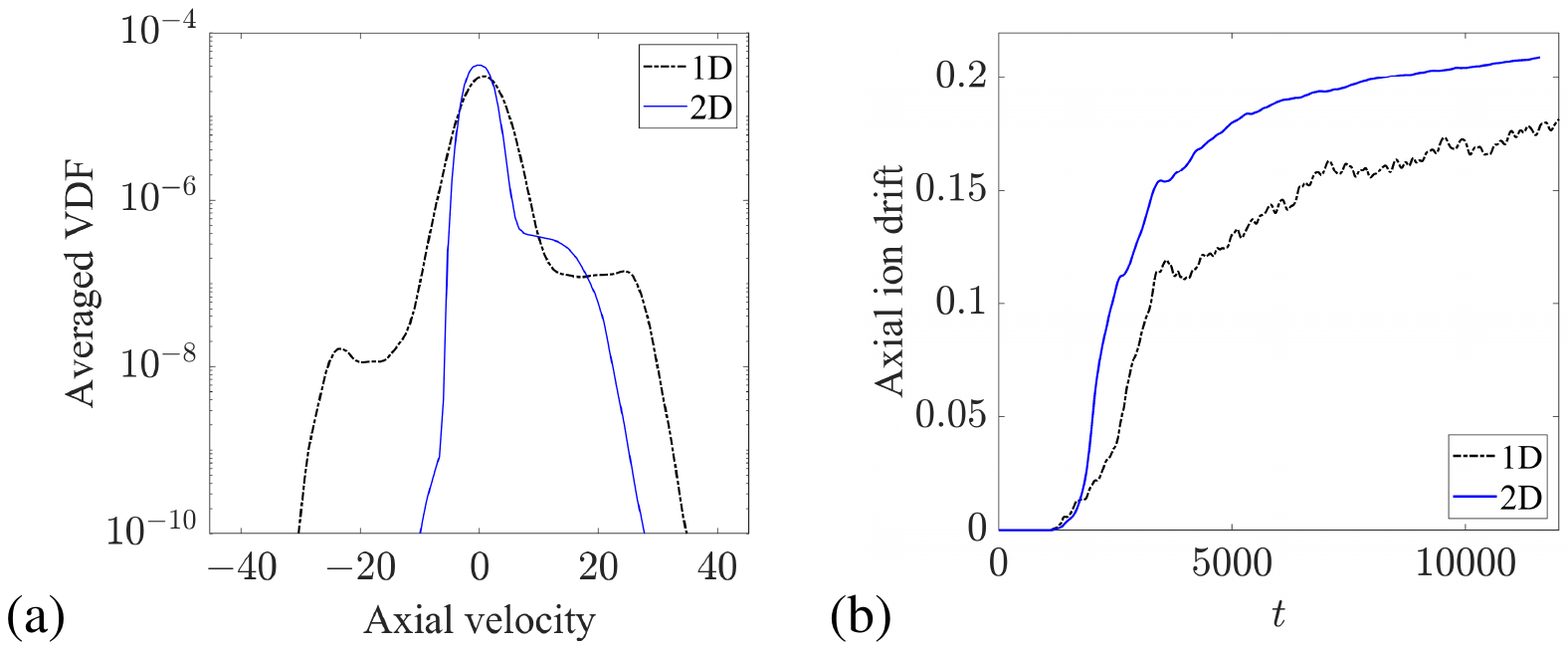}
      	\caption{(a) Comparison of 1D and 2D axial ion velocity distribution functions (VDFs), averaged over all points in space and $t \in [7\times10^3,9\times10^3]$ in the non-linear saturation regime, and integrated over all transverse velocities for the 2D VDF. (b) Comparison of time evolution of 1D and 2D axial ion drifts. Both plots are for $M_e = 2.8$. \label{fig:FVI1D2D}}
\end{center}
\end{figure}

Figure~\ref{fig:FVI1D2D}(a) plots a similar comparison of the averaged axial ion velocity distribution functions, $\ov{f}_i^{x,y,v_y}(v_x)$. Likewise, the 1D distribution extends to larger speeds as an indication of a larger effective ion temperature. In 1D, plateaus indicative of ion-acoustic wave behaviour are present at positive and negative axial velocities. A plateau is only observed in the positive axial direction in 2D. While the late-time ion velocity distribution functions in figure~\ref{fig:vdf2DtimeM2p3} suggested the presence of energetic backward-streaming ions in the 2D instability, these represent a smaller proportion of the ion population compared to those generated from a 1D instability. In other words, the multi-dimensional nature of the 2D instability appears to promote the earlier formation of transverse-streaming ions at the expense of backward-streaming ion formation due to significant early growth of transverse plasma waves (see also figures~\ref{fig:kx2D} and \ref{fig:ky2D}). Correspondingly, the mean axial ion velocity in the 1D simulations, where there are almost as many backward-streaming ions as forward-streaming ions, is smaller than that in the 2D simulations, where forward-streaming ions outnumber backward-streaming ions, as seen in figure~\ref{fig:FVI1D2D}(b). The greater symmetry of the ion distribution along the axial velocity direction in the 1D instability in figure~\ref{fig:FVI1D2D}(a), which corresponds to a less pronounced axial ion drift in figure~\ref{fig:FVI1D2D}(b), implies that the 1D plasma waves take on more of a standing than a travelling character in contrast to 2D axial waves as observed earlier, such as in the vortical structures of the left column of figure~\ref{fig:vdf2Dxvx}. This accounts for the more oscillatory nature of some of the 1D curves remarked earlier in the discussion of figure~\ref{fig:energetics1D2D}. 

\subsection{Potential amplitude and trapping frequency}

\begin{figure}
\begin{center}
    	\includegraphics[trim = 0 165 0 145, width = \textwidth]{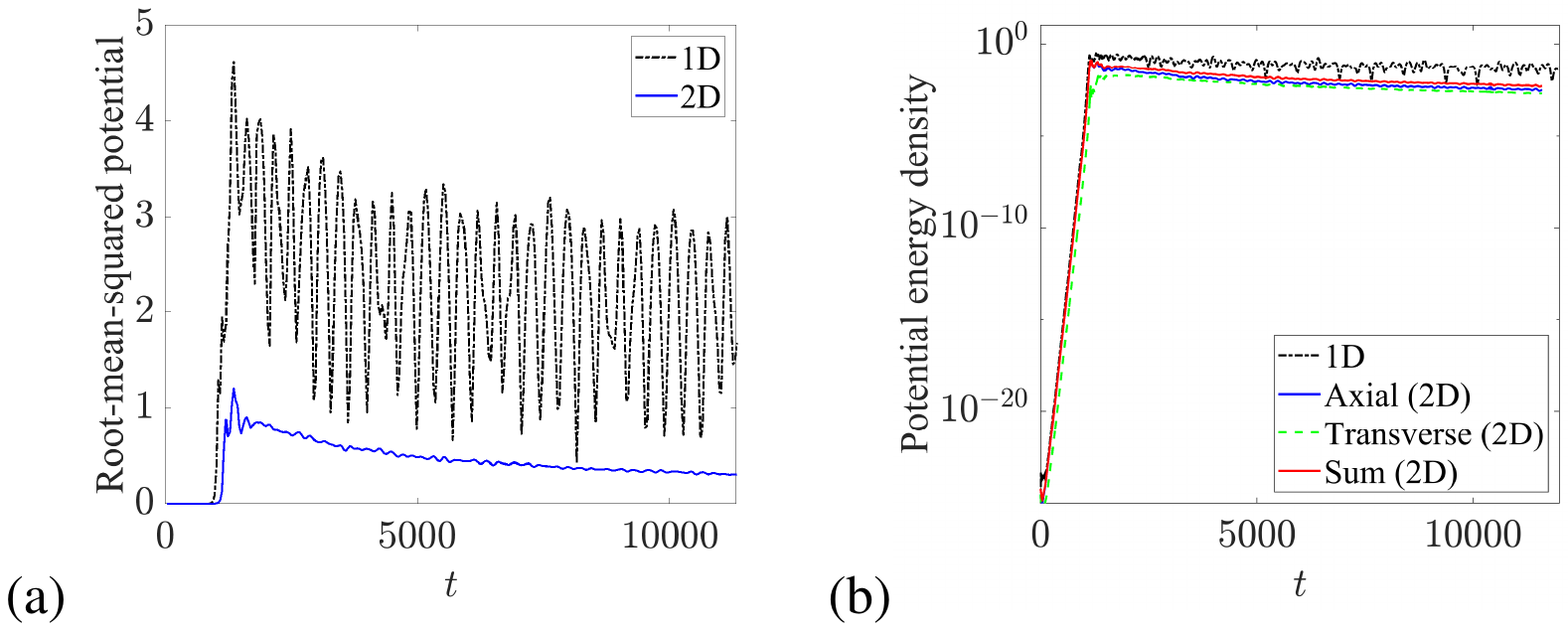}
      	\caption{(a) Comparison of time evolution of 1D and 2D root-mean-squared potentials. (b) Comparison of time evolution of 1D and 2D electrostatic potential energies. Both plots are for $M_e = 2.8$. \label{fig:electric1D2D}}
\end{center}
\end{figure}

We take a closer look at the effects of the electrostatic potential energy magnitude on ion and electron trapping. Trapping intensity may be quantified in more detail through the bounce frequency $g$, which is proportional to the square root of the potential amplitude, $\phi_\text{rms}$, keeping all else constant. Comparisons of the root-mean-squared potential are plotted in figure~\ref{fig:electric1D2D}(a). The 1D potential amplitude consistently exceeds its 2D counterpart by a factor of 4 or more, suggesting that trapping occurs at least twice as rapidly in the 1D instability than in the 2D one given that $g \sim \sqrt{\phi_\text{rms}}$. This may account for the lower degree of 2D energisation and the smaller tendency of the 2D instability to trap ions in the backward-streaming direction. As seen on a logarithmic scale in figure~\ref{fig:electric1D2D}(b), the difference in electrostatic potential energies between the 1D and 2D instabilities is of a couple of orders of magnitude, and not simply because energy is partitioned between the axial and transverse degrees of freedom. The smaller potential amplitudes and field strengths in the 2D system account for smaller effective ion and electron temperatures.

\section{Conclusions}\label{sec:conc}

\begin{figure}
\begin{center}
    	\includegraphics[trim = 0 0 0 0, width = \textwidth]{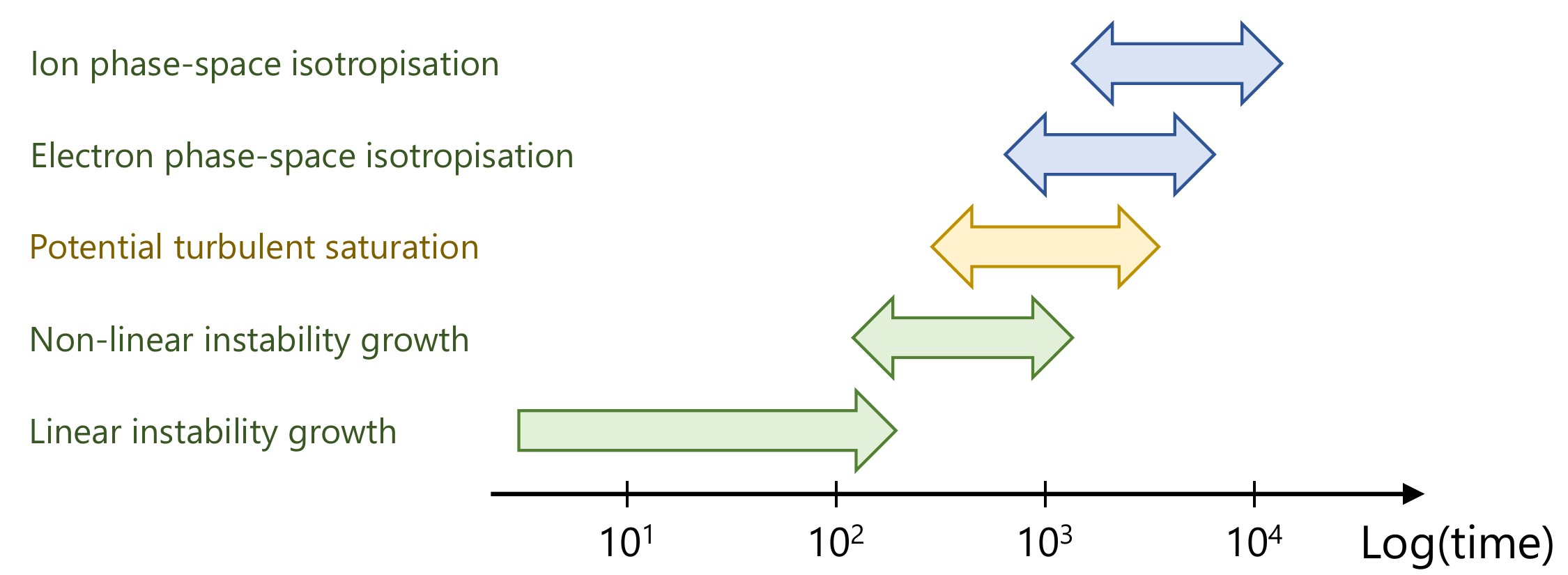}
      	\caption{Summary of instability growth and non-linear saturation process of various macroscopic and phase-space quantities as a function of the dimensionless time, $t$. Ion phase space isotropisation is arrested in the 2D instability.\label{fig:summary}}
\end{center}
\end{figure}

The investigation of electrostatic current-driven plasma instabilities is crucial to determine the origin and fluxes of high-energy ions. Such ions have practical implications including hollow cathode erosion in electric spacecraft thrusters, as well as the inception of magnetic reconnection and cosmic rays. A direct kinetic solver is used to study the evolution and directional dependence of the ion and electron velocity distribution functions, as well as their moments and accompanying electric potential, without contamination from statistical noise inherent in state-of-the-art particle methods. 

The electric potential field underlying such current-driven instabilities exhibits four developmental stages: linear modal growth, harmonic growth, accelerated growth via quasi-linear mechanisms alongside non-linear fill-in, and saturated turbulence. The maximum linear modal growth rate matches analytical predictions from the linear plasma dispersion relation. Harmonics of the fastest-growing fundamental, followed by intermediate-wavenumber modes, grow in a process that resembles the development of hydrodynamic turbulence. Ion and electron phase-space structures further exhibit a statistical tendency to isotropy also reminiscent of classical fluid behaviour. However, unlike hydrodynamic turbulence, which only fully emerges in 3D, such plasma turbulence occurs even in lower-dimensional instabilities as postulated by~\citet{buneman1959dissipation} since ions and electrons are allowed to interpenetrate unlike fluids. 

While the electrostatic potential energy quickly saturates and reaches a turbulent state, transverse-streaming and then backward-streaming ions are only formed after several ion trapping cycles. In other words, high-energy ions appear to be generated amid a turbulent potential field with statistically isotropic electrons. This formation process is accelerated and energised with a larger initial electron Mach number. A schematic summarising the instability growth and non-linear saturation process is shown in figure~\ref{fig:summary}.

The additional transverse degree of freedom in the 2D instability has several physical implications. A larger forward axial ion drift is observed due to a preference for transverse-streaming ions over backward-streaming ions, and the corresponding axial plasma waves are of a travelling rather than a standing nature. Steepening of a transient plateau occurs in the ion velocity distribution at the trapping speed as an indication of continued long-time plasma wave damping. The axial electron drift is smaller due to effective isotropisation of the electron distribution in the multi-dimensional velocity space. Despite the larger energy source resulting from this, smaller effective ion and electron temperatures are exhibited due to multi-dimensional energy redistribution, alongside smaller potential amplitudes and field strengths, which result in slower and weaker trapping as the system approaches a saturated turbulence state. On the whole, the degree of energisation in 2D appears to be lower than that in 1D, and the return to statistical isotropy of ion motion in phase space is correspondingly arrested, causing a less substantial presence of backward-streaming high-energy ions.

Under the warm-electron conditions studied in this work, it is shown that forward-streaming ions of up to 50 eV, as well as transverse-streaming and backward-streaming ions of up to 20 eV, are readily generated by electrostatic current-carrying instabilities. A detailed characterisation of such high-energy ion formation and directional distribution, which constitute precursors of the hollow cathode erosion process, can improve predictions of spacecraft thruster lifetime and complement accelerated life testing. A complete characterisation, however, necessitates the inclusion of additional physics such as three-dimensionality, collisions, and magnetisation, which will be addressed through further development and deployment of the direct kinetic method.

\section*{Acknowledgements}

The authors would like to acknowledge A.~R. Vazsonyi, J.~M. Wang, S.~S. Jain, S. Mirjalili, L. Jofre Cruanyes, K.~P. Griffin, and the multi-phase group at the 17th Stanford University Center for Turbulence Research Summer Program for helpful discussions and feedback. This work utilised the Blanca condo computing resource of the University of Colorado Boulder, as well as the Summit supercomputer, which is supported by the National Science Foundation (awards ACI-1532235 and ACI-1532236), the University of Colorado Boulder, and Colorado State University, and the Alpine high performance computing resource, which is jointly funded by the University of Colorado Boulder, the University of Colorado Anschutz, Colorado State University, and the National Science Foundation (award 2201538).

\appendix

\section{Grid convergence}\label{sec:gridconv}

\begin{figure}
\begin{center}
    	\includegraphics[trim = 0 165 0 145, width = \textwidth]{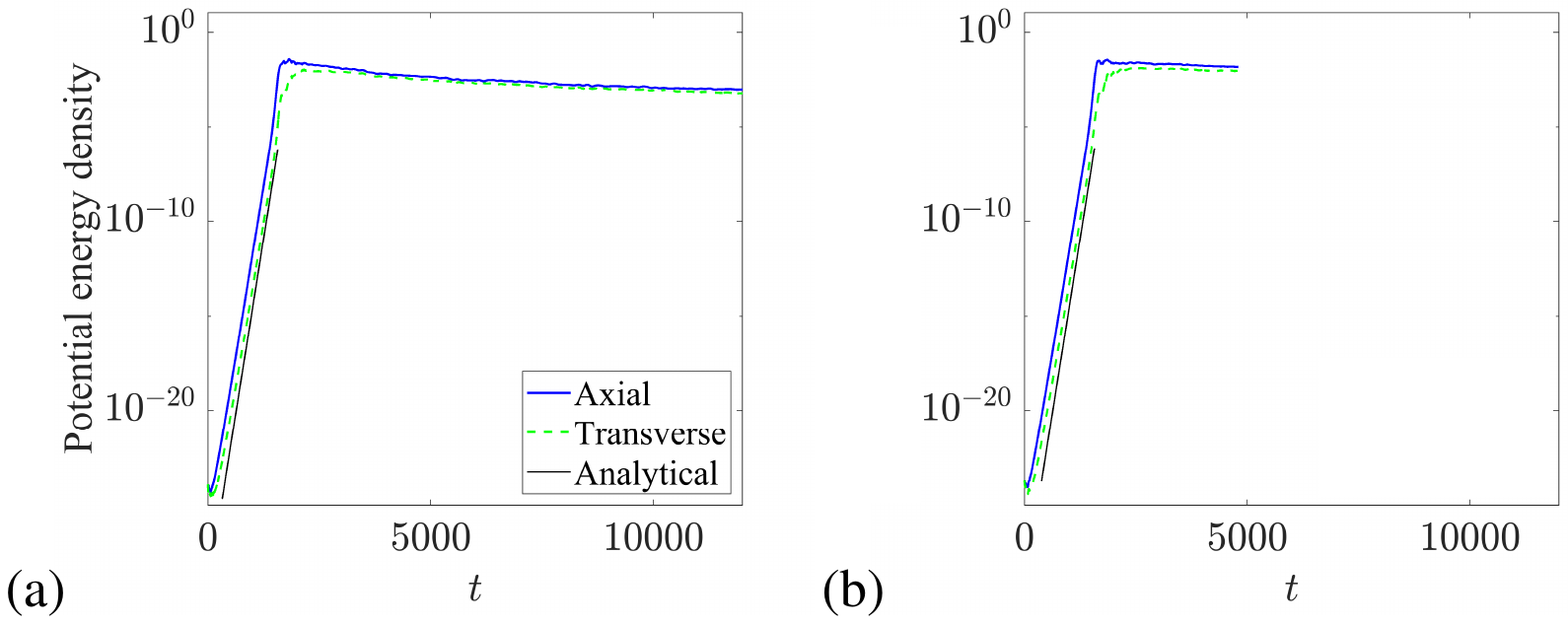}
      	\caption{Time evolution of the axial and transverse electrostatic potential energies for the (a) baseline and (b) velocity-refined simulations introduced in \S~\ref{sec:2D} with $M_e = 2.3$. The analytical lines denote the growth rate from linear stability theory via solution of \eqref{eqn:disp}. \label{fig:epe2D}}
\end{center}
\end{figure}

An additional simulation was performed with twice the resolution in all velocity dimensions relative to the baseline simulation to ascertain velocity grid convergence. Figure~\ref{fig:epe2D} plots the time evolution of the volume-averaged axial and transverse electrostatic potential energy densities, respectively, $\left(\int 0.5 E_x^2 \, \diff V\right)/L^2$ and $\left(\int 0.5 E_y^2 \, \diff V\right)/L^2$. Here, $E_x$ and $E_y$ are, respectively, the axial and transverse electric field strengths and the domain of integration is over the entire physical space. The baseline resolution exhibits satisfactory grid convergence up to the onset of saturation. The higher-resolution simulation is used in the analysis of Section~\ref{sec:stages}.

\bibliographystyle{jpp}

\bibliography{current-carrying}

\end{document}